\documentclass[a4paper,11pt]{article}
\pdfoutput=1 

\usepackage{jcappub}
\usepackage{graphicx}
\usepackage{dcolumn}
\usepackage{amssymb,amsmath,bm,bbold}
\usepackage{color}
\usepackage[dvipsnames,table]{xcolor}
\usepackage{xfrac}
\usepackage{aas_macros}
\usepackage{mathrsfs}
\usepackage{subcaption}
\usepackage{rotating}
\usepackage{chngcntr}
\usepackage{lipsum}
\usepackage{units}
\usepackage{multirow}
\usepackage{xspace}
\usepackage{xfrac}
\usepackage{mathtools}
\usepackage{comment}
\newcommand{\shear}{\boldsymbol{\gamma}}
\newcommand{\ellip}{\boldsymbol{e}}
\newcommand{\nv}{\hat{\boldsymbol{\theta}}}
\newcommand{\summ}[1]{\sum_{\bf #1}\Delta #1^2}
\newcommand{\Ylm}[3]{{\sf E}^{#1}_{\bf #2}({\bf #3})}
\newcommand{\Nside}{N_{\rm side}}
\newcommand{\cospar}{(\Omega_m, \allowbreak \Omega_b, \allowbreak h, \allowbreak n_s,\allowbreak \sigma_8)}

\newcommand{\dof}{\text{d.o.f.}}
\newcommand{\zbin}{$z$-bin\xspace}
\newcommand{\planck}{{\sl Planck}}
\newcommand{\mcal}{\textsc{Metacalibration}}
\newcommand{\nmt}{\texttt{NaMaster}}

\definecolor{internationalkleinblue}{rgb}{0.0, 0.18, 0.65}
\hypersetup{urlcolor=internationalkleinblue, linkcolor=internationalkleinblue, citecolor=internationalkleinblue}

\usepackage[T1]{fontenc} 
\usepackage{natbib}
\title{Cosmic shear power spectra in practice}

\author[a,1]{Andrina Nicola,}
\author[b,c,d]{Carlos Garc\'ia-Garc\'ia,}
\author[d]{David Alonso,}
\author[a,e]{Jo Dunkley,}
\author[d]{Pedro G. Ferreira,}
\author[f]{An\v ze Slosar,}
\author[a,g]{David N. Spergel}

\affiliation[a]{Department of Astrophysical Sciences, Princeton University, Peyton Hall, Princeton, NJ 08544, USA}
\affiliation[b]{Instituto de F\'isica Fundamental, Consejo Superior de Investigaciones Cient\'ificas, c/. Serrano 123, E-28006, Madrid, Spain}
\affiliation[c]{Institut de Ci\`{e}ncies del Cosmos (UB-IEEC), c/. Mart\'i i Franqu\'es 1, E-08028, Barcelona, Spain}
\affiliation[d]{Department of Physics, University of Oxford, Denys Wilkinson Building, Keble Road, Oxford OX1 3RH, United Kingdom}
\affiliation[e]{Department of Physics, Princeton University, Jadwin Hall, Princeton, NJ 08544, USA}
\affiliation[f]{Brookhaven National Laboratory, Physics Department, Upton, NY 11973, USA}
\affiliation[g]{Center for Computational Astrophysics, Flatiron Institute, 162 Fifth Avenue, New York, NY 10010, USA}

\emailAdd{anicola@astro.princeton.edu}

\abstract{Cosmic shear is one of the most powerful probes of Dark Energy, targeted by several current and future galaxy surveys. Lensing shear, however, is only sampled at the positions of galaxies with measured shapes in the catalog, making its associated sky window function one of the most complicated amongst all projected cosmological probes of inhomogeneities, as well as giving rise to inhomogeneous noise.
Partly for this reason, cosmic shear analyses have been mostly carried out in real-space, making use of correlation functions, as opposed to Fourier-space power spectra. Since the use of power spectra can yield complementary information and has numerical advantages over real-space pipelines, it is important to develop a complete formalism describing the standard unbiased power spectrum estimators as well as their associated uncertainties. Building on previous work, this paper contains a study of the main complications associated with estimating and interpreting shear power spectra, and presents fast and accurate methods to estimate two key quantities needed for their practical usage: the noise bias and the Gaussian covariance matrix, fully accounting for survey geometry, with some of these results also applicable to other cosmological probes.
We demonstrate the performance of these methods by applying them to the latest public data releases of the Hyper Suprime-Cam and the Dark Energy Survey collaborations, quantifying the presence of systematics in our measurements and the validity of the covariance matrix estimate. We make the resulting power spectra, covariance matrices, null tests and all associated data necessary for a full cosmological analysis publicly available.}

\begin{document}
\maketitle
\flushbottom

  \section{Introduction}\label{sec:intro}
    Since its first detection two decades ago \cite{astro-ph/0003008, astro-ph/0003338, astro-ph/0002500, astro-ph/0003014}, cosmic shear has become a powerful cosmological probe of the late-time Universe, uniquely sensitive to its Dark Matter content and the properties of Dark Energy \cite{astro-ph/0609591, 1201.2434}. It therefore lies at the core of several current and future surveys, including the Dark Energy Survey (DES)\footnote{\url{https://www.darkenergysurvey.org}.}, the Hyper Suprime-Cam survey (HSC)\footnote{\url{https://hsc.mtk.nao.ac.jp/ssp}.} and the Kilo-Degree Survey (KiDS)\footnote{\url{http://kids.strw.leidenuniv.nl}.}, as well as the Rubin Observatory Legacy Survey of Space and Time (LSST)\footnote{\url{https://www.lsst.org}.}, Euclid\footnote{\url{https://www.euclid-ec.org}.} and the Roman Telescope\footnote{\url{https://roman.gsfc.nasa.gov}.}.

    Cosmic shear measurements are obtained from the shapes of individual galaxies and the shear field can therefore only be reconstructed at discrete galaxy positions, making its associated angular masks some of the most complicated amongst those of projected cosmological observables. This is in addition to the usual complexity of large-scale structure masks due to the presence of stars and other small-scale contaminants. So far, cosmic shear has therefore mostly been analyzed in real-space as opposed to Fourier-space (see e.g. Refs.~\cite{1708.01538, 2007.15633}, while Ref.~\cite{1809.09148} used power spectra). However, Fourier-space analyses offer complementary information and cross-checks as well as several advantages, such as simpler covariance matrices, and the possibility to apply simple, interpretable scale cuts. Therefore, methodologies to perform direct Fourier-space analyses for cosmic shear have been studied in the literature \cite{1004.3542, 1603.07818}. In addition, several indirect cosmic shear power spectrum estimation methods \cite{astro-ph/0206182, 1412.3851} or closer cousins such as COSEBIs \cite{1002.2136} have been proposed. Common to these methods is that power spectra are derived by Fourier transforming real-space correlation functions, thus avoiding the challenges pertaining to direct approaches.

    Given the fast growth of available imaging data in current and future experiments, and the presence of statistical tensions between the parameter constraints from some of these datasets and those from Cosmic Microwave Background (CMB) experiments \cite{1708.01530,1809.09148,1906.09262,2007.15632}, it is vital to have a tight control over the level of accuracy of the estimators used to calculate the different ingredients of the likelihood. For instance, the weak lensing community have identified the accuracy of the covariance matrix \cite{1202.2332,1606.05338,1905.06454}, including the effects of survey geometry on the shape noise contributions \cite{1804.10663}, and the choice of angular scales at which to evaluate the theoretical predictions for binned correlation functions \cite{1707.06627,1804.10663}, as details that can have a significant effect on the final parameter constraints. As we will discuss here, these problems can be addressed accurately and analytically through the use of power spectra. This is in addition to the larger problems of photometric redshift uncertainties \cite{1602.05960,1906.09262,1909.09632,2004.09542}, modeling the effects of baryons on the matter power spectrum \cite{1111.0052,1407.4301,1904.07905,1905.06082}, or the impact of intrinsic galaxy alignments \cite{astro-ph/0005269,astro-ph/0005384,1305.5791,1303.1808}, which we do not address here.

    In this work, we build on Refs.~\cite{1004.3542, 1603.07818} and present methodologies for efficient and accurate cosmic shear analyses in Fourier-space, especially focusing on two challenges faced by these methods: the estimation of the noise power spectrum, or noise bias due to intrinsic galaxy shape noise and the estimation of the Gaussian contribution to the power spectrum covariance. We present analytic expressions for both the shape noise contribution to cosmic shear auto-power spectra and the Gaussian covariance matrix, which fully account for the effects of complex survey geometries. These expressions avoid the need for potentially expensive simulation-based estimation of these quantities. In addition, even though the cosmic shear covariance receives additional contributions from super-sample variance (e.g. \cite{1302.6994}) and the non-Gaussian connected trispectrum (e.g. \cite{astro-ph/0012087}), cosmic shear errors have been found to be dominated by the Gaussian covariance \cite{1807.04266}, making its accurate estimation an important part of any weak lensing analysis.
    
    We validate the derived expressions by applying them to the latest public data releases of two current cosmic shear surveys, HSC Data Release 1 \cite{1702.08449} and DES Year 1 \cite{1801.03181,1708.01531}. We perform a rigorous suite of systematics tests and compare our results to those obtained by the respective collaborations \cite{1708.01538, 1809.09148}. Furthermore, we make all the products needed to use these data for cosmological analyses publicly available at \url{https://github.com/xC-ell/ShearCl}.
    
    This paper is organized as follows. In Section \ref{sec:pcl}, we briefly review pseudo-$C_\ell$ power spectrum estimation and derive  analytic expressions for both noise bias and Gaussian covariance matrices within this framework. In Section \ref{sec:data}, we present the data sets used in this work and the validation of our results using these data is presented in Section \ref{sec:results}. We conclude in Section \ref{sec:conclusion}. Appendix \ref{app:pix} discusses the effective pixel window function in cosmic shear datasets, and Appendix \ref{app:nulls} contains further details on the null tests performed.
    
  \section{Pseudo-$C_\ell$s for cosmic shear}\label{sec:pcl}
    \begin{table*}
      \centering
      \begin{tabular}{|c|c|c|}
        \hline
        Symbol & Curved sky & Flat sky (continuum $\rightarrow$ discretized) \\
        \hline
        \(\displaystyle {\bf l} \) & \(\displaystyle (\ell,m) \) & \(\displaystyle (l_x,l_y) \) \\
        \(\displaystyle \summ{l} \) & \(\displaystyle \sum_{\ell=0}^{\infty} \sum_{m=-\ell}^{\ell} \) & \(\displaystyle \int \frac{dl^2}{2\pi} \rightarrow \sum_{\bf l}\frac{2\pi}{L_xL_y}\) \\
        \(\displaystyle \Delta^x({\bf l},{\bf l}')\) & \(\displaystyle \delta^K_{\ell\ell'}\delta^K_{mm'}\) & \(\displaystyle 2\pi\,\delta^D({\bf l}-{\bf l}')\rightarrow \delta^K_{l_xl'_x}\delta^K_{l_yl'_y}\frac{L_xL_y}{(2\pi)^2}2\pi\) \\
        \(\displaystyle {\bf x}\) & \(\displaystyle \nv\equiv(\theta,\varphi) \) & \(\displaystyle (x,y)\) \\
        \(\displaystyle \summ{x} \) & \(\displaystyle \int_0^{\phi} d\phi \int_{-1}^{1} d(\cos\theta)\) & \(\displaystyle \int \frac{dx^2}{2\pi}\rightarrow\sum_{\bf x}\frac{L_xL_y}{2\pi N_xN_y}\) \\
        \(\displaystyle \Delta^k({\bf x},{\bf y})\) & \(\displaystyle \delta^D(\cos\theta-\cos\theta')\delta^D(\varphi-\varphi')\) & \(\displaystyle 2\pi\,\delta^D({\bf x}-{\bf y})\rightarrow\delta^K_{xx'}\delta^K_{yy'}\frac{N_xN_y}{L_xL_y}2\pi\) \\
        \(\displaystyle V \) & \(\displaystyle 1 \) & \(\displaystyle (2\pi)^{-1}\) \\
        \hline
      \end{tabular}
      \caption{Lookup table describing the generalized notation introduced in Ref.~\cite{1906.11765} for quantities defined on the sphere (second column) and on the flat 2D plane (third column). For the flat-sky case, we also provide expressions for a discretized, finite 2D plane with periodic boundary conditions. In this case, the map has dimensions $(L_x,L_y)$ subdivided into $(N_x,N_y)$ equi-spaced pixels in $(x,y)$. $\delta^D$ and $\delta^K$ denote the Dirac and Kronecker delta functions, respectively.}\label{tab:notation}
    \end{table*}

    This section summarizes the mathematical framework behind pseudo-$C_\ell$ estimators. In particular, we will focus on the problems of estimating the noise bias and disconnected covariance matrix in the presence of a complex mask, describing general methods to calculate both accurately. We will first briefly describe cosmic shear and its measurement in order to give a specific example for the generation of the fields considered in this work. The subsequent sections, describing power spectrum estimation, employ a generic notation applicable to the analysis of any projected field.

    \subsection{Cosmic shear}\label{ssec:pcl.shear}
    
        The lensing shear $\shear\equiv(\gamma_1,\gamma_2)$ is a spin-2 field that is, at linear order, proportional to the traceless components (the ellipticity $\ellip\equiv(e_1,e_2)$) of the projected inertia tensor of an originally circular source \cite{astro-ph/9912508}. Cosmic shear can be thus estimated from the measured ellipticities of galaxy images, but the presence of a finite point spread function and noise in the images conspire to complicate its unbiased measurement. Modern shear estimation methods can generally be subdivided into three classes: point estimators (e.g. \cite{astro-ph/9411005}), the \mcal{}/\textsc{Metadetect} shear estimation method \cite{1702.02600, 1911.02505} and the Bayesian Fourier domain (BFD) approach \cite{1304.1843}. All of these methods apply different corrections for the measurement biases arising in cosmic shear. We refer the reader to the respective papers and Sections \ref{ssec:data.hsc} and \ref{ssec:data.des} for more details. 
        
        In the simplest model, the measured shear of a single galaxy can be decomposed into the actual shear, a contribution from measurement noise and the intrinsic ellipticity of the galaxy. Intrinsic galaxy ellipticities dominate the observed shears and single object shear measurements are therefore noise-dominated. Moreover, intrinsic ellipticities are correlated between neighboring galaxies or with the large-scale tidal fields, leading to correlations not caused by lensing, usually called ``intrinsic alignments''. For the purposes of this discussion, we will absorb all correlated components into the signal $\shear$ and assume uncorrelated noise. With this subdivision, the intrinsic alignment signal must be modeled as part of the theory prediction for cosmic shear. Finally we note that measured shears are prone to leakages due to the point spread function ellipticity and its associated errors. These sources of contamination must be either kept at a negligible level, or modeled and marginalized out.
        
        To compute power spectra, the catalogs obtained from galaxy surveys must be pixelized onto a finite grid, assigning each pixel the weighted average of the ellipticities of all galaxies falling into it:
        \begin{equation}\label{eq:shear_map}
        \hat{\shear}({\bf x}_p)=\frac{\sum_{i\in p}v_i\,\ellip_i}{\sum_{i\in p} v_i},
        \end{equation}
        where $\shear_i$ denotes the shear of the $i$-th galaxy, $v_i$\footnote{Shape weights are usually labelled $w$, but we use $v$ here to avoid confusing it with the weight map $w({\bf x}_p)$, introduced in the subsequent sections.} is its shear measurement weight and sums are over all galaxies falling into pixel $p$. Analogously, we can compute the per-pixel noise variance map from the galaxy ellipticities as
        \begin{equation}\label{eq:ellip_var}
        \sigma_{\gamma,p}^2=\frac{\sum_{i\in p} v_i^2\sigma_{e,i}^2}{\left(\sum_{i\in p}v_i\right)^2},\hspace{12pt}
        \sigma_{e,i}^2\equiv\frac{1}{2}(e_{1,i}^2+e_{2,i}^2).
      \end{equation}
      We note that this expression is equivalent to the noise variance that would result from averaging over a large suite of random catalogs in which the original ellipticities of all sources are rotated by independent random angles. Ellipticities rotated by an angle $\alpha$, $e_k^\alpha$, are related to the un-rotated ones via a double-angle phase: $e_1^\alpha+ie^\alpha_2=\exp(i2\alpha)\,(e_1+ie_2)$. The shear variance over many such rotations can be calculated by averaging $(e_1^\alpha)^2$ or $(e_2^\alpha)^2$ over $\alpha$ with a flat distribution in the interval $[0,2\pi)$, yielding ${\rm Var}(e_k^\alpha)=(e_1^2+e_2^2)/2$, as in Eq. \ref{eq:ellip_var}. Using this expression in all equations below is therefore equivalent to estimating the corresponding quantities (noise bias, covariance matrices) by computing power spectra from a large number of randomly rotated catalogs, as done in e.g. Refs.~\cite{1507.05598,1809.09148}. As these random rotations destroy any existing correlation between galaxy ellipticities, their use will not capture any source of correlated noise. Generating random catalogs, their corresponding shear maps and their power spectra are computationally expensive operations for large datasets. Therefore, the methods presented in this work significantly simplify the estimation of power spectra and their covariance matrices while simultaneously improving the accuracy of those estimates by eliminating the stochastic noise associated with a finite number of random realizations.

In addition to the signal and noise maps discussed above, we also need their associated sky masks or weight maps in order to compute power spectra. The inverse variance weight map for shear observations for a general set of weights $v_i$ and shear estimate variances $\sigma_{e,i}$ is
\begin{equation}\label{eq:weight_map_IV}
      w_{\rm IV}({\bf x}_p) = \frac{\left(\sum_{i\in p}v_i\right)^2}{\sum_{i\in p} v_i^2\sigma_{e,i}^2}.
\end{equation}
In the case of inverse variance weighting, which amounts to setting $v_i = \sigma_{e,i}^{-2}$, Eq.~\ref{eq:weight_map_IV} reduces to\footnote{This is an optimal weighting when $\sigma_{e,i}$ contains all sources of variance of shear around its mean value in the pixel, including that coming from intrinsic shear fluctuations on small scales. In practice these are always subdominant to the measurement error.}
        \begin{equation}\label{eq:weight_map_SW}
          w_{\rm SW}({\bf x}_p) = \sum_{i\in p}v_i.
        \end{equation}

      Eq.~\ref{eq:shear_map} implies that shears can only be estimated in pixels in which galaxies have been detected, thus giving rise to complex sky masks. In addition, the noise in pixelized cosmic shear maps depends on the number of galaxies falling into each pixel and is therefore inhomogeneous, as can be seen from Eq.~\ref{eq:ellip_var}. This is immediately more complicated than the case of e.g. galaxy clustering, where the noise is homogeneous for a uniform survey (up to depth variations). These reasons, as well as challenges arising from the spin-2 nature of the involved fields, render the measurement of power spectra for cosmic shear more involved than for other probes.
      
      Finally, the use of pixelized datasets requires accounting for the effects of the finite pixel size. Assuming that the value contained in each pixel is a homogeneous \emph{average} of a given field, this is commonly accounted for by deconvolving the pixel window function in Fourier space (see e.g. \cite{1004.3542}), alongside any other sources of smoothing (e.g. instrumental beams). If instead the pixels contain the values of the field \emph{sampled} at a given point (e.g. the pixel centre), no smoothing needs to be accounted for. Since, again, the cosmic shear signal is only sampled at discrete source positions, finite-pixel effects cannot be modeled exactly, since they depend on the number density and distribution of those sources, as well as pixel size. For low number densities ($\sim1$ source per pixel on average), the situation is closer to the \emph{sampling} case, while for high densities ($\sim10$ sources per pixel), pixelization is more akin to \emph{averaging}. This is discussed in more detail in Appendix \ref{app:pix}. The general advice, nevertheless, is to use a pixel size $\theta_{\rm pix}$ that is smaller than the smallest scale used in the analysis (i.e.  $\ell_{\rm max}\ll\pi/\theta_{\rm pix}$), and to test that the results are insensitive to the choice of pixel resolution, as is done in this work.

    \subsection{Preliminaries}\label{ssec:pcl.prelim}
    
      To derive results that are simultaneously applicable in curved skies and within the flat-sky approximation, we will use the abstract notation introduced in Ref.~\cite{1906.11765} and summarized in Tab.~\ref{tab:notation}. Further details about this formalism can be found in Refs.~\cite{1809.09603,1906.11765}.

      The data considered here will be in the form of a pixelized 2D map of a Gaussian, statistically-isotropic, stochastic field ${\bf a}({\bf x})$ with 1 or 2 components per pixel (for spin-0 and spin-2 quantities, respectively).
      Its generalized Fourier coefficients are given by \begin{equation}
        {\bf a}_{\bf k}=\summ{x}\,\Ylm{\dag}{k}{x}\,{\bf a}({\bf x}),
      \end{equation}
      where the operator $\summ{x}$ denotes an integral or sum over all values of the coordinates ${\bf x}$, and $\Ylm{\dag}{k}{x}$ are a complete set of orthogonal functions given in terms of spin-weighted spherical harmonics or 2D plane waves as described in Ref.~\cite{1809.09603}. The orthogonality and completeness relations are:
      \begin{align}
        &\summ{x}\,\Ylm{\dag}{k}{x}\Ylm{}{l}{x}=\mathbb{1}\Delta^x({\bf k},{\bf l}),\\
        &\summ{k}\,\Ylm{}{k}{x}\Ylm{\dag}{k}{y}=\mathbb{1}\Delta^k({\bf x},{\bf y}),
      \end{align}
      where $\Delta^x$ and $\Delta^k$ are generalized delta functions in Fourier and real space respectively and $\mathbb{1}$ denotes the identity matrix in the spin matrix basis.
      The statistics of these Gaussian fields are determined by their covariance, defined in terms of the power spectrum ${\sf C}^{ab}_k$ as
      \begin{equation}\label{eq:iso}
        \langle {\bf a}_{\bf k}{\bf b}^\dag_{\bf l}\rangle\equiv {\sf C}^{ab}_\ell\,V\,\Delta^x({\bf k},{\bf l}),
      \end{equation}
      where $V$ is a volume factor. 

    All maps will be composed of signal (${\bf s}$) and noise (${\bf n}$), so that ${\bf a}({\bf x})={\bf s}^a({\bf x})+{\bf n}^a({\bf x})$, where the superscript $a$ is added to specify that we refer to the signal and noise of the field $a$. Using a discrete notation, where ${\bf x}_p$ is the position of the sky pixel with index $p$, we will assume that the noise is uncorrelated between pixels, but not necessarily homogeneous. Therefore we have
      \begin{equation}
        \langle {\bf n}^a({\bf x}_p)\,{\bf n}^{b\dagger}({\bf x}_{p'})\rangle = (\sigma^{a}_p)^2\mathbb{1}\delta_{ab}\delta_{pp'},
      \end{equation}
      where we have further assumed that the noise components of different maps are uncorrelated. If this noise is uncorrelated on arbitrarily small scales, the noise variance should scale inversely with the pixel area $\Omega_{\rm pix}$ (i.e. $(\sigma_p^a)^2\propto \Omega_{\rm pix}^{-1}$).

    \subsection{Pseudo-$C_\ell$}
      In the presence of inhomogeneous noise, the data must be optimally weighted in order to minimize the variance of the estimated power spectrum. Optimal quadratic estimators use a well-educated guess of the data's covariance matrix and multiply the data by this inverse covariance (e.g. \citep{astro-ph/9611174, astro-ph/9708203}). While several methods have been proposed to speed up this calculation (e.g. Refs.~\cite{astro-ph/9805339, astro-ph/0104443, astro-ph/0304513, astro-ph/0209560}), this remains a computationally challenging task for high-resolution data. Pseudo-$C_\ell$ estimators circumvent this problem by assuming a diagonal covariance, in which case inverse-variance weighting is a simple product of maps, where the weighted or "observed" map is \citep{astro-ph/0105302}:
      \begin{equation}
        \tilde{\bf a}({\bf x}_p)=w^a({\bf x}_p)\,{\bf a}({\bf x}_p)=w^a({\bf x}_p)\,\left[{\bf s}^a({\bf x}_p)+{\bf n}^a({\bf x}_p)\right].
      \end{equation}
      If the map is noise-dominated, and assuming uncorrelated noise as described above, the optimal weight map $w^a({\bf x})$ (also commonly called the ``mask'') is simply
      \begin{equation}\label{eq:invar}
        w^a({\bf x}_p) = \frac{N^a}{(\sigma^a_p)^2},
      \end{equation}
      where $N^a$ is an arbitrary constant that could be chosen to e.g. make $w^a$ unitless\footnote{For signal-dominated maps, a more appropriate weighting scheme would be found by adding the signal and noise variances in quadrature. The optimality of the pseudo-$C_\ell$ estimator in this case, however, would depend on the steepness of the signal power spectrum \citep{1306.0005}.
      }.

      Given a choice of weight $w^a$, the Fourier coefficients of $\tilde{\bf a}$ are
      \begin{align}
        \tilde{\bf a}_{\bf l}&=\sum_{\bf k}(\Delta k)^2\,{\sf M}_{{\bf l}{\bf k}}(w^a){\bf s}^a_{\bf k}+\sum_{\bf p}({\Delta}x)^2w^a({\bf x}_p){\sf E}^\dagger_{\bf l}({\bf x}_p){\bf n}^a({\bf x}_p).
      \end{align}
      The quantities ${\sf M}_{{\bf l}{\bf k}}(w)$ are the coupling coefficients, given by
      \begin{equation}
        {\sf M}_{{\bf l}{\bf k}}(w)\equiv\sum_{p}(\Delta x)^2\,w({\bf x}_p){\sf E}^\dagger_{\bf l}({\bf x}_p){\sf E}_{\bf k}({\bf x}_p),
      \end{equation}
      and we have left the noise part expressed in real space, where its covariance is simplest.
 
      The diagonal of the coupling coefficients ${\sf M}_{{\bf l}{\bf k}}$ satisfies:
      \begin{align}\label{eq:coup_diag}
        &{\sf M}_{{\bf l}{\bf l}}(w) = \mathbb{1}\frac{A}{2\pi}\langle w\rangle_{\rm pix},\hspace{12pt}\,\text{in flat sky},\\
        &\frac{\sum_{m=-\ell}^\ell{\sf M}_{{\bf l}{\bf l}}(w)}{2\ell+1} = \mathbb{1}\langle w\rangle_{\rm pix},\hspace{12pt}\,\text{in curved sky},
      \end{align}
      where $\langle\cdot\rangle_{\rm pix}$ denotes averaging the quantity inside brackets over all pixels, and $A$ is the patch area.

      Since signal and noise are uncorrelated, the covariance of two observed fields $\tilde{\bf a}_{\bf l}$ and $\tilde{\bf b}_{\bf l}$ is:
      \begin{align}\label{eq:pc2pt}
        \langle \tilde{\bf a}_{\bf l}\tilde{\bf b}^\dagger_{{\bf l}'}\rangle=&V\sum_{\bf k} (\Delta k)^2 {\sf M}_{{\bf l}{\bf k}}(w^a){\sf C}^{ab}_k{\sf M}^\dagger_{{\bf l}'{\bf k}}(w^b)\,+\delta_{ab}(\Delta x)^2{\sf M}_{{\bf l}{\bf l}'}((\sigma^aw^a)^2).
      \end{align}
      Given this result, the pseudo-$C_\ell$ estimator proceeds in two steps:
      \begin{enumerate}
        \item We first group different ${\bf l}$ modes into bins (typically bands of similar $\ell$ or annuli of flat-sky Fourier modes spanning a range of radii). The resulting binned pseudo-$C_\ell$'s are called bandpowers and are given by:
        \begin{equation}\label{eq:bandpowers}
          \tilde{\sf C}^{ab}_q=\sum_{{\bf l}\in q} B_q^{\bf l}\,\tilde{\bf a}_{\bf l}\tilde{\bf b}^\dag_{\bf l},
        \end{equation}
        where the index $q$ denotes a given bandpower and its weights are normalized such that $\sum_{{\bf l}\in q}B_q^{\bf l}=(V\Delta^x({\bf 0}))^{-1}$.
        
        Using Eq. \ref{eq:pc2pt}, the ensemble average of the binned pseudo-spectrum is related to the true underlying power spectrum through a linear operation:
        \begin{equation}
          \left\langle \tilde{\sf C}^{ab}_q\right\rangle=\sum_{{\bf l}\in q} B_q^{\bf l}V\sum_{\bf k} (\Delta k)^2 {\sf M}_{{\bf l}{\bf k}}(w^a){\sf C}^{ab}_k{\sf M}^\dagger_{{\bf l}{\bf k}}(w^b)+\delta_{ab}\tilde{\sf N}_q,
        \end{equation}
        where $\tilde{\sf N}_q$ collects the second term in Eq. \ref{eq:pc2pt} and will be discussed in the next section. 
        \item The correlation between different elements of the pseudo-spectrum induced by the mode-coupling coefficients is largely reversed by multiplying $\tilde{\sf C}^{ab}_q$ by the inverse of the so-called binned ``mode-coupling matrix'' $\mathcal{M}$, which is related to the mode-coupling coefficients ${\sf M}_{{\bf l}{\bf k}}$ and the explicit expression can be found in Eq. (14) in Ref.~\cite{1809.09603}. With this, the final pseudo-$C_\ell$ estimator is given by
        \begin{equation}\label{eq:decouple}
          {\rm vec}\left(\hat{\sf C}^{ab}_q\right) = \sum_{q'}\left(\mathcal{M}^{-1}\right)_{qq'}\,{\rm vec}\left(\tilde{\sf C}^{ab}_{q'}-\delta_{ab}\tilde{\sf N}_{q'}\right).
        \end{equation}
        Here the ${\rm vec}()$ operator transforms an $N\times N$ matrix into an $N^2$ vector by concatenating all its rows and hat denotes the final uncoupled estimator of the power spectrum.
        
        The main numerical advantage of pseudo-$C_\ell$ estimators is the fact that $\mathcal{M}_{qq'}$ can be computed analytically, making use of methods scaling as $\ell_{\rm max}^3$ \citep{astro-ph/0105302}.
      \end{enumerate}
      This procedure is fully analytic, allowing us to account for the effects of mode coupling and binning on the theory prediction exactly. The binned theory prediction ${\sf C}^{ab}_q$ is given by the unbinned power spectrum ${\sf C}^{ab}_\ell$ convolved with the so-called {\sl bandpower window function} ${\cal F}_{q\ell}$. In the case of spin-0 fields defined on a curved sky, this is given by \citep{1809.09603}:
      \begin{equation}\label{eq:bpwkernel}
        C^{ab}_q = \sum_\ell {\cal F}^{ab}_{q\ell}\,C^{ab}_\ell,\hspace{12pt}
        {\cal F}^{ab}_{q\ell}\equiv\sum_{q'}\left(\mathcal{M}^{-1}\right)_{qq'}\sum_{\ell'\in q'}(2\ell+1)\,\Xi_{\ell'\ell}(w^a,w^b),
      \end{equation}
      where we have defined the coupling coefficients
      \begin{equation}\label{eq:coupXi}
        \Xi_{\ell\ell'}(w^a,w^b)=\frac{\sum_{m=-\ell}^\ell\sum_{m'=-\ell'}^{\ell'}M_{{\bf l}{\bf l}'}(w^a)M^*_{{\bf l}{\bf l}'}(w^b)}{(2\ell+1)(2\ell'+1)}.
      \end{equation}
      For spin-2 fields, the bandpower window functions additionally involve mixing of $E$ and $B$-modes and the specific expressions for the binned theory power spectra can be found in Ref.~\cite{1809.09603}. There is therefore no ambiguity in the procedure to calculate a binned theory power spectrum: the theory prediction evaluated at all integer multipoles needs to be convolved with the linear kernel ${\cal F}^{ab}_{q\ell}$ in Eq. \ref{eq:bpwkernel}. In real-space analyses, the equivalent procedure involves integrating the correlation function over bins of angular separation weighted by the number of weighted pairs of sources at different angles. This operation can be prohibitively expensive, but may be circumvented by evaluating the correlation function at judiciously chosen scales \citep{1804.10663,1810.02353}.
      
    \subsection{Noise bias}\label{ssec:pcl.noise}
    The second term in Eq. \ref{eq:pc2pt} is the so-called noise bias, which must be subtracted from auto-correlations in Eq. \ref{eq:decouple} in order to obtain an un-biased estimate of the signal power spectrum. Using the property of the mode-coupling coefficients in Eq. \ref{eq:coup_diag}, the contribution to the binned pseudo-spectrum is white (i.e. scale-independent) and given by:
      \begin{equation}\label{eq:noise_bias}
        \tilde{\sf N}_q=\left\langle\sum_{{\bf l}\in q} B_q^{\bf l} |\tilde{\bf n}_{\bf l}|^2\right\rangle=\mathbb{1}\Omega_{\rm pix}\left\langle w^2\,\sigma^2\right\rangle_{\rm pix},
      \end{equation}
      where $\Omega_{\rm pix}$ is the pixel area in steradians, which we assume to be constant across the map. Note that this result is valid in both curved and flat skies. The contribution to the mode-decoupled bandpowers is then computed by inverting the binned mode-coupling matrix:
      \begin{equation}
        \hat{\sf N}_q = \sum_{q'}\left(\mathcal{M}^{-1}\right)_{qq'}\tilde{\sf N}_{q'}.
      \end{equation}
      These expressions show that under the assumptions made in this work, any scale dependence in the final mode-decoupled noise bias is purely induced by the mode-coupling matrix.
      
      For the specific case of shear, the noise bias to the pseudo-spectrum (Eq. \ref{eq:noise_bias}) is given by:
        \begin{equation}\label{eq:noise_bias_shear}
          \tilde{\sf N}^{\rm SW}_\ell=\mathbb{1}\Omega_{\rm pix}\left\langle \sum_{i\in p} v_i^2\sigma_{e,i}^2\right\rangle_{\rm pix},
          \hspace{12pt}
          \tilde{\sf N}^{\rm IV}_\ell=\mathbb{1}\Omega_{\rm pix}\left\langle \frac{\left(\sum_{i\in p}v_i\right)^2}{\sum_{i\in p} v_i^2\sigma_{e,i}^2}\right\rangle_{\rm pix}\,,
        \end{equation}
        for sum of weights (SW) or inverse variance (IV) masks, respectively.

    \subsection{Covariances}\label{ssec:pcl.covar}
      Accurately calculating the covariance of the estimated power spectra is crucial in order to use them to extract reliable cosmological parameter constraints. Due to the non-linear nature of gravitational collapse, the initially Gaussian density fluctuations become non-Gaussian over time. While the distribution of the pseudo-$C_\ell$ estimator on sufficiently small scales is still well described as a multivariate Gaussian \cite{0801.0554,1707.04488} due to the central limit theorem, the covariance of this distribution receives contributions from the connected (i.e. non-Gaussian) four-point correlator of the density field. These additional contributions have been extensively studied in the literature (e.g. \cite{astro-ph/0012087,0810.4170,0906.2237,1302.6994}), and can be broken down into two components: the so-called super-sample covariance (SSC), caused by the non-linear coupling of modes larger than the surveyed volume with small-scale fluctuations, and the standard non-Gaussian connected trispectrum (cNG). Ref.~\cite{1807.04266} found that the disconnected or ``Gaussian'' part of the covariance matrix (i.e. the covariance matrix calculated under the assumption that all fields involved are Gaussianly distributed), constitutes the dominant source of uncertainty, and that the SSC dominates over the cNG terms for cosmic shear on most scales.
      
      Since the Gaussian covariance matrix dominates the final uncertainties, its accurate analytical estimation will be the focus of our discussion in this section. The results presented in Section \ref{ssec:results.release} and the data products made publicly available will also include the SSC and cNG contributions, estimated analytically following the approach of Ref.~\cite{1601.05779}\footnote{As the cosmic shear covariance is dominated by the Gaussian part, we model the effects of finite sky coverage on the non-Gaussian corrections using the approximations given in Ref.~\cite{1601.05779} and do not fully account for mode-coupling as we do for the Gaussian part.}. Analytic estimates of the covariance matrix are computationally less expensive than simulation-based approaches and have recently been shown to yield consistent constraints, even in the case of a full-shape galaxy power spectrum analysis \cite{2009.00622}.

      \subsubsection{Gaussian covariances}\label{sssec:pcl.covar.gaussian}
        The main complication in estimating the disconnected covariance is accounting for the mode-coupling effects induced by incomplete sky coverage. Survey geometry has been identified as a key factor in obtaining accurate uncertainties for cosmic shear data \citep{1804.10663} which, if incorrectly modeled, can lead to wrong assessments of goodness-of-fit, parameter biases and varying levels of tension between experiments.
        
        The disconnected covariance of the pseudo-$C_\ell$ estimator was studied in detail in Ref.~\cite{1906.11765} in the context of large-scale structure data, and similar approaches have been studied in Refs.~\cite{astro-ph/0307515,astro-ph/0410394,1609.09730,1811.05714}. Here we briefly describe the main aspects of the formalism described in Ref.~\cite{1906.11765}, and extend it to account for the interplay between noise and signal, as well as the increased level of mode-coupling caused by the cosmic shear mask. To simplify the discussion and notation, we will describe the main aspects of Gaussian pseudo-$C_\ell$ covariances in the context of spin-0 fields with a single component. The main results shown here are then directly applicable to fields with arbitrary spins following the approximations described in Ref.~\cite{1906.11765}.

        The pseudo-$C_\ell$ estimator involves products of two fields, and its covariance is therefore a linear combination of the four-point function of those fields. Under the assumption of Gaussian statistics, we can use Wick's theorem to express those four-point functions in terms of products of two-point correlators. As shown in Ref.~\cite{astro-ph/0307515}, estimating the covariance between the power spectrum $\hat{\sf C}^{ab}_\ell$ of fields $a$ and $b$, and that of fields $f$ and $g$, $\hat{\sf C}^{fg}_{\ell'}$, reduces to computing terms of the form:
        \begin{equation}
          \langle \tilde{a}_{\bf l}\tilde{g}^*_{{\bf l}'}\rangle\langle \tilde{b}^*_{\bf l}\tilde{f}_{{\bf l}'}\rangle+(f\leftrightarrow g),
        \end{equation}
        where the second term involves interchanging the roles of fields $f$ and $g$.
        
        Each of these terms involves a two-point correlator of the masked fields as in Eq. \ref{eq:pc2pt} at different Fourier modes ${\bf l}$ and ${\bf l}'$. Computing those exactly for ${\bf l}\neq{\bf l}'$ is a slow $O(\ell^6_{\rm max})$ calculation, since the usual mathematical identities that simplify the calculation of the pseudo-$C_\ell$ mode-coupling matrix cannot be applied. For that reason the so-called ``narrow kernel approximation'' (NKA) has been advocated in Refs.~\cite{astro-ph/0307515,1906.11765}. This approximation is based on assuming that the coupling coefficients $M_{{\bf l}{\bf k}}$ have a narrow peak around ${\bf l}={\bf k}$, and that all power spectra $C^{ab}_k$ vary slowly within that region (and can therefore be taken out of the sum over ${\bf k}$ modes). The standard NKA reads:
        \begin{equation}\label{eq:nka}
          \langle \tilde{a}_{\bf l}\tilde{b}^*_{{\bf l}'}\rangle=\sum_{\bf k}(\Delta k)^2 M_{{\bf l}{\bf k}}(w^a)M_{{\bf l'}{\bf k}}^*(w^b)C^{ab}_k\simeq C^{ab}_{(\ell,\ell')}\sum_{\bf k}(\Delta k)^2 M_{{\bf l}{\bf k}}(w^a)M_{{\bf l'}{\bf k}}^*(w^b),
        \end{equation}
        where $C^{ab}_{(\ell,\ell')}=\sfrac{(C^{ab}_\ell+C^{ab}_{\ell'})}{2}$. This is further simplified due to the following property of the coupling coefficients:
        \begin{equation}\label{eq:coupling_product}
          \sum_{\bf k}(\Delta k)^2 M_{{\bf l}{\bf k}}(w^a)M_{{\bf l'}{\bf k}}^*(w^b) = M_{{\bf l}{\bf l}'}(w^aw^b).
        \end{equation}
        Thus, in the NKA, Eq. \ref{eq:pc2pt} reads:
        \begin{align}\label{eq:pc2pt_nka}
          \langle \tilde{a}_{\bf l}\tilde{b}^*_{{\bf l}'}\rangle=&VC^{ab}_{(\ell,\ell')}M_{{\bf l}{\bf l}'}(w^aw^b) + \delta_{ab}(\Delta x)^2M_{{\bf l}{\bf l}'}((\sigma^aw^a)^2).
        \end{align}
        After averaging over annuli, the covariance between the pseudo-spectra becomes:
        \begin{align}\nonumber
          {\rm Cov}(\tilde{C}^{ab}_\ell,\tilde{C}^{fg}_{\ell'})=
          &C^{ag}_{(\ell,\ell')}C^{bf}_{(\ell,\ell')} \Xi_{\ell\ell'}(w^aw^g,w^bw^f)\,+\\\nonumber
          &\delta_{ag}(\Delta x)^2 C^{bf}_{(\ell,\ell')} \Xi_{\ell\ell'}((w^a\sigma^a)^2,w^bw^f)\,+\\\nonumber
          &\delta_{bf}(\Delta x)^2 C^{ag}_{(\ell,\ell')} \Xi_{\ell\ell'}(w^aw^g,(w^b\sigma^b)^2)\,+\\\nonumber
          &\delta_{ag}\delta_{bf}(\Delta x)^4 \Xi_{\ell\ell'}((w^a\sigma^a)^2,(w^b\sigma^b)^2)\\\label{eq:covar_pcl}
          &+(f\leftrightarrow g),
        \end{align}
        where we have used the coupling coefficients defined in Eq. \ref{eq:coupXi}. Note that the arguments of the coupling coefficients $\Xi_{\ell,\ell'}$ are products of masks and noise variance maps, instead of individual maps as in Eq. \ref{eq:coupXi}.
      
        Due to the usage of the NKA, Eq. \ref{eq:covar_pcl} is approximate in the signal terms, but it is exact in the noise terms. A similar result was presented in Ref.~\cite{1811.05714}. In addition, within this approximation, the mode-coupling matrices used to estimate the pseudo-$C_\ell$ between two fields with masks $w^a$ and $w^b$ are given by
        \begin{equation}     
          \mathcal{M}_{\ell\ell'}=(2\ell'+1)\,\Xi_{\ell\ell'}(w^a,w^b),
        \end{equation}
        i.e. the coupling coefficients $\Xi_{\ell\ell'}$ are equivalent (up to a factor of $2\ell'+1$) to the usual mode-coupling matrices, except they involve other types of maps. Thus, within the NKA, estimating the Gaussian pseudo-$C_\ell$ covariance is roughly as fast as estimating the power spectra themselves. In principle, however, it requires calculating four additional sets of mode-coupling coefficients, which usually dominates the total computation time.

      \subsubsection{Approximating the noise contribution}\label{sssec:pcl.covar.noise}
        Although Eq. \ref{eq:covar_pcl} is in principle no more complicated than the noiseless case, having to calculate and store three extra mode-coupling matrices can be an unnecessary drag on computational resources. It is therefore worth exploring whether an adequately chosen effective ``noise power spectrum'' ${\sf N}_\ell^{\rm cov}$ could be added to the signal $C_\ell$ in the first term of the equation to replace the other three. To do so, let us again turn to Eq. \ref{eq:pc2pt_nka}. In the case ${\bf l}={\bf l}'$, $a=b$, after averaging over annuli, and using the property of the coupling coefficients (Eq. \ref{eq:coup_diag}), we obtain:
        \begin{equation}
          \left\langle |\tilde{a}_{\bf l}|^2\right\rangle_{|{\bf l}|}=\langle (w^a)^2\rangle_{\rm pix}\left[C^{ab}_\ell+\Omega_{\rm pix}\frac{\langle (w^a\sigma^a)^2\rangle_{\rm pix}}{\langle (w^a)^2\rangle_{\rm pix}}\right],
        \end{equation}
        where $\langle\cdot\rangle_{|{\bf l}|}$ implies averaging over annuli. Thus, a natural approximation to the noise power spectrum to be used for covariance calculation is:
        \begin{equation}\label{eq:noicov}
          {\sf N}_\ell^{\rm cov}=\mathbb{1}\Omega_{\rm pix}\frac{\langle w^2 \sigma^2\rangle_{\rm pix}}{\langle w^2\rangle_{\rm pix}}=\mathbb{1}\Omega_{\rm pix}\frac{\langle \sigma^{-2}\rangle_{\rm pix}}{\langle \sigma^{-4}\rangle_{\rm pix}},
        \end{equation}
        where the second equality holds for inverse-variance weights (Eq. \ref{eq:invar}). Note that this ``covariance'' noise spectrum is the same as the mode-coupled noise bias $\tilde{\sf N}_q$ in Eq. \ref{eq:noise_bias}, corrected for the impact of mode coupling by a simple rescaling factor $1/\langle w^2\rangle_{\rm pix}$. 

        In the case of shear, the approximate noise spectrum to be used for covariance estimation (Eq. \ref{eq:noicov}) is given by
        \begin{align}
          &{\sf N}_\ell^{\rm cov,SW}=\mathbb{1}\Omega_{\rm pix}\frac{\left\langle \sum_{i\in p}v_i^2\sigma_{e,i}^2\right\rangle_{\rm pix}}{\left\langle \left(\sum_{i\in p}v_i\right)^2\right\rangle_{\rm pix}}, \label{eq:noicov_shear1}
          \\
          &{\sf N}_\ell^{\rm cov, IV}=\left\langle \frac{\left(\sum_{i\in p}v_i\right)^2}{\sum_{i\in p} v_i^2\sigma_{e,i}^2}\right\rangle_{\rm pix}\bigg/\left\langle \frac{\left(\sum_{i\in p}v_i\right)^4}{\left(\sum_{i\in p} v_i^2\sigma_{e,i}^2\right)^2}\right\rangle_{\rm pix},\label{eq:noicov_shear2}
        \end{align}
      for sum of weights or inverse variance masks, respectively. This approximation will be tested in Section~\ref{ssec:results.method.covar}.

      \subsubsection{Improving the NKA}\label{sssec:pcl.covar.nkaplus}
        The NKA is likely to break down for sufficiently complicated weight maps where the coupling coefficients $M_{{\bf l}{\bf k}}$ have support over a broad range of ${\bf k}$ around ${\bf l}$. If the width of this range, $\delta k$, is larger than the typical scale of variation of the power spectrum $\delta k_{C_\ell}\sim (d\log C_\ell/d\ell)^{-1}$, the power spectrum on the left hand side of Eq.~\ref{eq:nka} will be effectively smoothed over the support of $M_{{\bf l}{\bf k}}$, which will cause it to differ significantly from the original $C_\ell$. The NKA can therefore be improved by replacing the symmetrized spectrum on the right hand side of Eq. \ref{eq:nka} with its smoothed version. This can be done through the following approximation
        \begin{equation}\label{eq:nka_plus}
           \sum_{\bf k}(\Delta k)^2M^a_{{\bf l}{\bf k}}M^{b*}_{{\bf l}'{\bf k}}C^{ab}_k\simeq \left.\frac{\left\langle\sum_{\bf k}(\Delta k)^2M^a_{{\bf l}{\bf k}}M^{b*}_{{\bf l}{\bf k}}C^{ab}_k\right\rangle_{|{\bf l}|}}{\left\langle\sum_{\bf k}(\Delta k)^2M^a_{{\bf l}{\bf k}}M^{b*}_{{\bf l}{\bf k}}\right\rangle_{|{\bf l}|}}\right|_{(\ell,\ell')}\sum_{\bf k}(\Delta k)^2M^a_{{\bf l}{\bf k}}M^{b*}_{{\bf l}'{\bf k}},
        \end{equation}
        where we have used the shorthand $M^a_{{\bf l}{\bf k}}\equiv M_{{\bf l}{\bf k}}(w^a)$.
        
        A close inspection of the location of the ${\bf l}$ and ${\bf l}'$ indices in Eq. \ref{eq:nka_plus} shows that the basis of this approximation is to calculate the diagonal elements ${\bf l}={\bf l}'$ exactly, and to approximate the structure of the  off-diagonal elements as a rescaling of the mode-coupling product in Eq. \ref{eq:coupling_product}. The difference of this approach with respect to the standard NKA is that we have not yet discarded the effect that mode coupling has on the $C_\ell$ (i.e. we haven't taken the $C_k$ out of the sum in the numerator), but we have assumed that the relative correlations between different ${\bf l}$ and ${\bf l}'$ are well captured by the convolution of coupling coefficients.
        
        Hence, this improvement over the original NKA results simply in substituting $C^{ab}_\ell$ in Eq. \ref{eq:covar_pcl} by the {\sl mode-coupled} pseudo-$C_\ell$
        \begin{align}\nonumber
          {\cal C}^{ab}_\ell &\equiv \frac{\left\langle\sum_{\bf k}(\Delta k)^2M^a_{{\bf l}{\bf k}}M^{b*}_{{\bf l}{\bf k}}C^{ab}_k\right\rangle_{|{\bf l}|}}{\left\langle\sum_{\bf k}(\Delta k)^2M^a_{{\bf l}{\bf k}}M^{b*}_{{\bf l}{\bf k}}\right\rangle_{|{\bf l}|}}\\\label{eq:sigcov}
          &=\frac{\langle \tilde{C}^{ab}_{\ell}\rangle}{\langle w^aw^b\rangle_{\rm pix}},
        \end{align}
        where we have used the definition of the pseudo-spectrum (Eq. \ref{eq:bandpowers}), together with the properties of the mode-coupling matrices (Equations \ref{eq:coup_diag} and \ref{eq:coupling_product}).
        
        In analogy with the result found in the previous section for the noise contribution (Eq. \ref{eq:noicov}), the signal power spectrum entering the covariance matrix is the mode-coupled signal power spectrum corrected for the effects of mode coupling by a simple rescaling factor $1/\langle w^aw^b\rangle_{\rm pix}$. This simple rescaling explains why the $1/f_{\rm sky}$ type pseudo-$C_\ell$ estimators (see e.g. Refs.~\cite{1973ApJ...185..413P,astro-ph/0105302}) work: Although these estimators are biased in general, this bias is small in the case of almost white power spectra or simple sky masks. Thus, in this limit, the improved NKA presented here reduces to the original NKA.

  \section{Data}\label{sec:data}
    As a practical demonstration of the methods described in the previous section, we apply them to the latest public data releases from the HSC and DES collaborations. We describe these data here.

    \subsection{HSC DR1}\label{ssec:data.hsc}
      The Hyper Suprime-Cam Subaru Strategic Program (HSC SSP) is an ongoing, wide-field imaging survey, which is currently in its last year of operation. The survey is expected to cover approximately 1000 deg$^{2}$ in the $grizy$ broadband photometric bandpasses.
    
      In this work, we use data from the first public data release of the HSC SSP (HSC DR1)\footnote{\url{https://hsc-release.mtk.nao.ac.jp/doc/}.}, which cover an area of approximately 130 deg$^2$ and consist of three different layers: Wide, Deep and UltraDeep \cite{1702.08449}. We use data from the Wide layer, which is subdivided into six different fields and covers approximately 108 deg$^2$ to a limiting magnitude of $m_{\mathrm{lim}, i} \sim 26.4$ with median $i$-band seeing of around 0.6 arcsec.
    
      We select galaxies suitable for cosmic shear measurement by applying the selection cuts given in Ref.~\cite{1705.06745}. Photometric galaxy redshifts in HSC are estimated using six different algorithms. Following Ref.~\cite{1809.09148}, we use the photo-$z$ estimate from \texttt{Ephor\_AB} to further subdivide these data into four tomographic redshift bins with edges $[0.3, 0.6, 0.9, 1.2, 1.5]$. This selection leaves us with $2{,}642{,}125$, $2{,}588{,}321$, $1{,}927{,}581$ and $1{,}109{,}056$ galaxies per redshift bin, respectively.
 
      The publicly available HSC galaxy shapes are measured from coadded $i$-band images using the re-Gaussianization method presented in Ref.~\cite{astro-ph/0301054}, and are corrected for PSF effects. The measured galaxy ellipticities $\hat{\ellip}$ must additionally be corrected for possible shape measurement biases. For HSC, the calibrated galaxy ellipticities are given by \cite{1705.06745}:
      \begin{equation}\label{eq:hsc:ecal}
        \ellip_i=\frac{1}{1+\bar{m}}\left(\frac{\hat{\ellip}_{i}}{2 \mathcal{R}}-\boldsymbol{c}_i\right),
      \end{equation}
      where $\boldsymbol{c}_{i}$ and $\bar{m}$ are the so-called \emph{additive} and \emph{multiplicative} biases and $\mathcal{R}$ denotes the shear responsivity, which quantifies the response of the measured ellipticity to an external shear \cite{astro-ph/9411005, astro-ph/0107431}. In HSC, these shear measurement biases are estimated using image simulations \cite{1710.00885}. Each galaxy in the weak lensing sample carries a weight, which is based on the inverse of the sum of shear measurement and shape noise, and we use these weights to construct cosmic shear maps. Following Ref.~\cite{1705.06745}, we further employ these weights to estimate the shear responsivity $\mathcal{R}$ and the weighted average of the multiplicative bias $\bar{m}$ used for shear calibration in each HSC field and tomographic redshift bin separately. The additive bias correction on the other hand, is applied on a per-object level.
   
      In order to measure photometric redshift distributions for each of the tomographic bins considered in this analysis, we follow the method presented in Ref.~\cite{1912.08209}, which is based on the COSMOS 30-band catalog presented in Ref.~\cite{1604.02350}.

    \subsection{DES Y1 data}\label{ssec:data.des}
       The Dark Energy Survey (DES) is a completed imaging survey, covering $\unit[5000]{deg^2}$ in 5 filter bands $(g,r,i,z,Y)$. The $10\sigma$ limiting magnitudes of the galaxy sample are $g = 23.4$, $r = 23.2$, $i = 22.5$, $z = 21.8$ and $Y = 20.1$. In this work, we use publicly available data from the DES Year 1 (Y1) data release, which  covers $\unit[1786]{deg^2}$ after coaddition and before masking ~\cite{1801.03181,1708.01531}\footnote{The data are available at \url{https://des.ncsa.illinois.edu/releases/y1a1}.}. 

      We follow Ref.~\cite{1708.01533} and use the so-called source galaxy sample for cosmic shear power spectrum measurements. In DES Y1, galaxy shapes are
      determined using two different algorithms: \mcal{}~\cite{1702.02600,1702.02601} and \textsc{IM3SHAPE}~\cite{1708.01533}. \mcal{}~\cite{1702.02600,1702.02601} fits a 2D Gaussian model for each galaxy to the pixel data in the $r$, $i$ and $z$ exposures, convolved with the corresponding point-spread function (PSF). In order to calibrate the shear estimator, this process is repeated with artificially sheared images in order to determine the response of the estimator, and to calibrate shear-dependent selection effects (see e.g. Refs.~\cite{1708.01533,1708.01537,1708.01538,1702.02601}). \textsc{IM3SHAPE}~\cite{1302.0183}, instead, maximizes the likelihood of a bulge-plus-disk model for each galaxy's $r$-band image. This estimator is biased and calibrated through image simulations~\cite{1708.01533,1708.01534}. In this work, we restrict our analysis to the \mcal{} catalog and only include galaxies with declination ${\rm dec.}<-35^\circ$, as was done in Ref.~\cite{1708.01533}. We further subdivide this sample into the same four tomographic bins used in Ref.~\cite{1708.01533}, containing $7{,}705{,}486$, $7{,}851{,}711$, $8{,}238{,}547$, and $4{,}196{,}641$ objects, in order of increasing mean tomographic bin redshift. The resulting sample covers an area of $\sim1320\,{\rm deg}^2$.

      Each of these four tomographic redshift bins has an associated redshift distribution $n_i(z)$. In DES Y1, galaxies are subdivided into redshift bins based on their photo-$z$ posterior as derived by the {\tt BPZ} algorithm \cite{astro-ph/9811189}. Further details on the construction and calibration of these redshift distributions can be found in Ref.~\cite{1708.01532}. We note that the determination of the redshift distributions in both the HSC and DES samples used in this work relies on COSMOS~\cite{1604.02350}. This inevitably correlates the parameter constraints found by both datasets, which makes the combination of their data non-trivial.

      For \mcal{}, the calibrated ellipticities are given by 
      \begin{equation}\label{eq:des:ecal}
        \ellip_i=\hat{\ellip}_i/\bar{R},
      \end{equation}
      where the additive bias is zero by construction and the multiplicative bias is given by $\bar{R}=({\sf R}_{11}+{\sf R}_{22})/2$. The quantity ${\sf R}={\sf R}_{\shear}+{\sf R}_{\rm S}$ denotes the $2\times 2$ response matrix. It accounts for the shear response, ${\sf R}_{\shear}$, which measures the change in ellipticity under a shear $\shear$ and the selection response, ${\sf R}_{\rm S}$, originating from biases induced by applying selection criteria on the sheared galaxy sample. The response matrix was found to be well approximated by a diagonal with ${\sf R}_{11} \approx {\sf R}_{22}$. As the estimate of ${\sf R}_{\shear}$ is noisy, this quantity is averaged over all sources in each redshift bin before using it in Eq.~\ref{eq:des:ecal}. This is common practice in shear power spectrum analyses \citep{1507.05598,1509.04071,1809.09148}, but differs from the approach used in real-space studies, where the multiplicative correction is applied at the two-point function level. Alternatively, in order to account for possible spatial variations in the shear response, one can compute local averages of ${\sf R}_\gamma$ over sufficiently large patches \citep{1903.04957}. Finally, we note that Ref.~\cite{1708.01538} observed a significant mean ellipticity in the sample ($\bar{\ellip}_i\sim\mathcal{O}(10^{-4})$) of unclear origin. We follow Ref.~\cite{1708.01538} and subtract these biases from all galaxies in each redshift bin.

  \section{Results}\label{sec:results}
     We now apply the methods described in Section \ref{sec:pcl} to the HSC and DES data described above. First, we briefly describe the data analysis pipeline used to estimate the cosmic shear power spectra from the public catalogs (Sec.~\ref{ssec:results.pipe}). We then use simulations to validate our calculation of the noise bias and covariance matrix (Sec.~\ref{ssec:results.method}), before presenting the estimated power spectra and their validation in Sec.~\ref{ssec:results.val}. These power spectra are made publicly available, and the released data is briefly described in Sec.~\ref{ssec:results.release}.
     \begin{figure}
        \centering
        \includegraphics[width=0.8\textwidth]{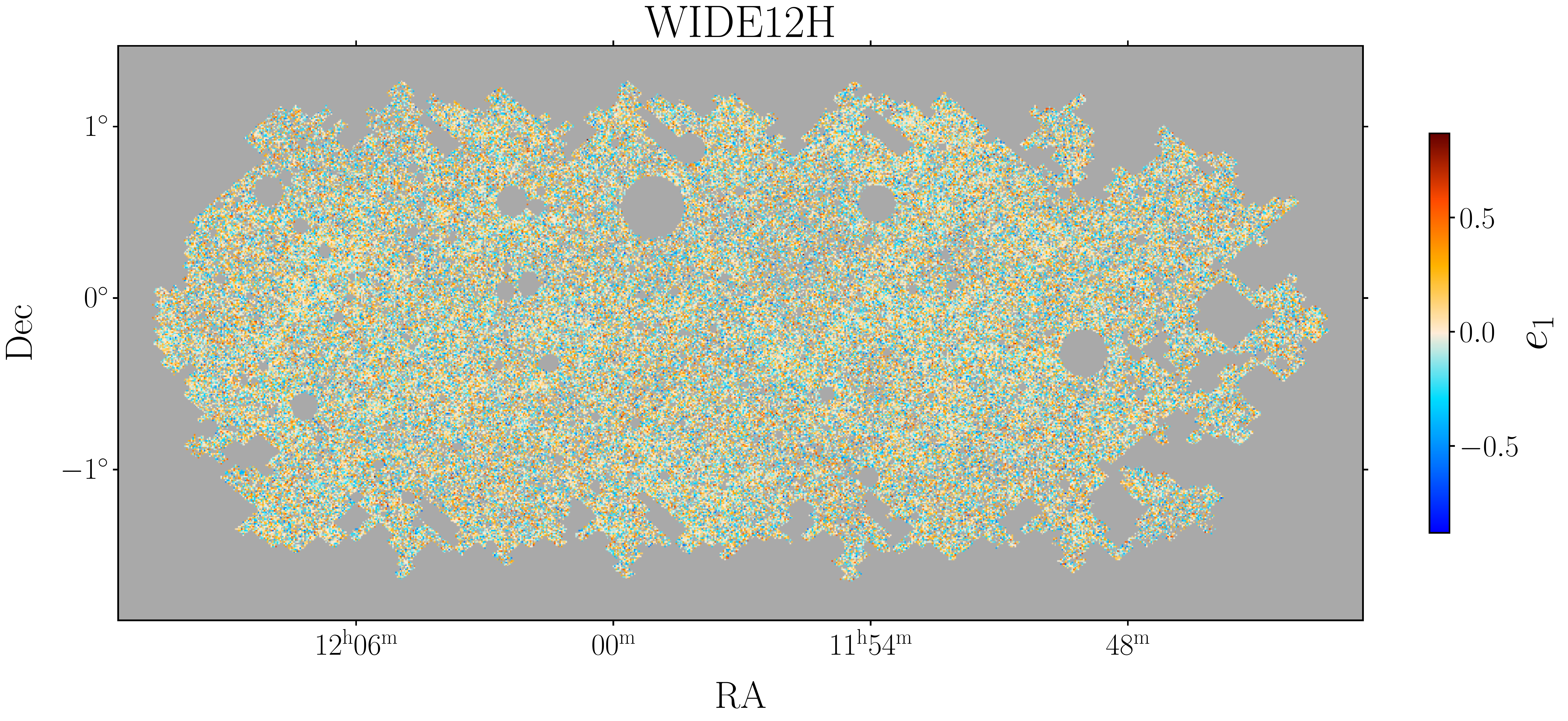}
        \includegraphics[width=0.8\textwidth]{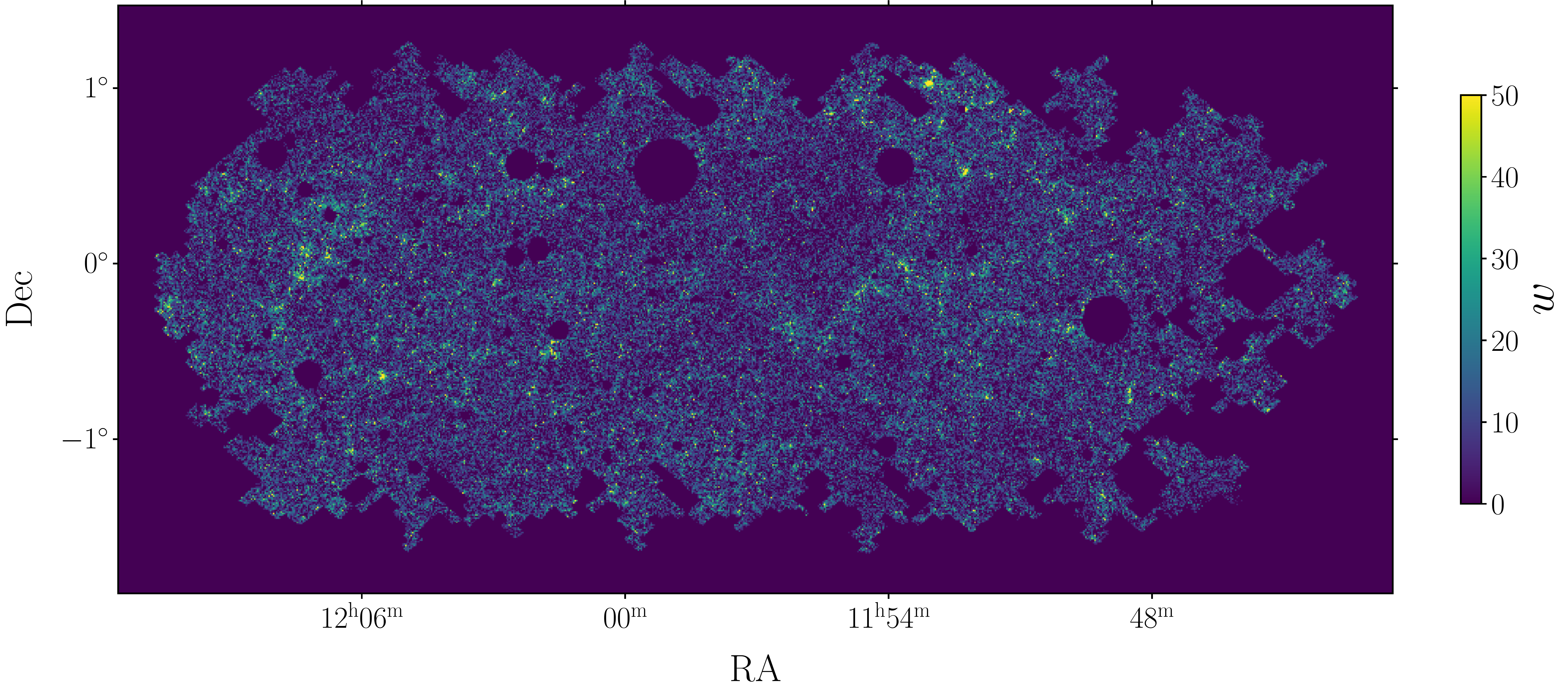}
        \includegraphics[width=0.8\textwidth]{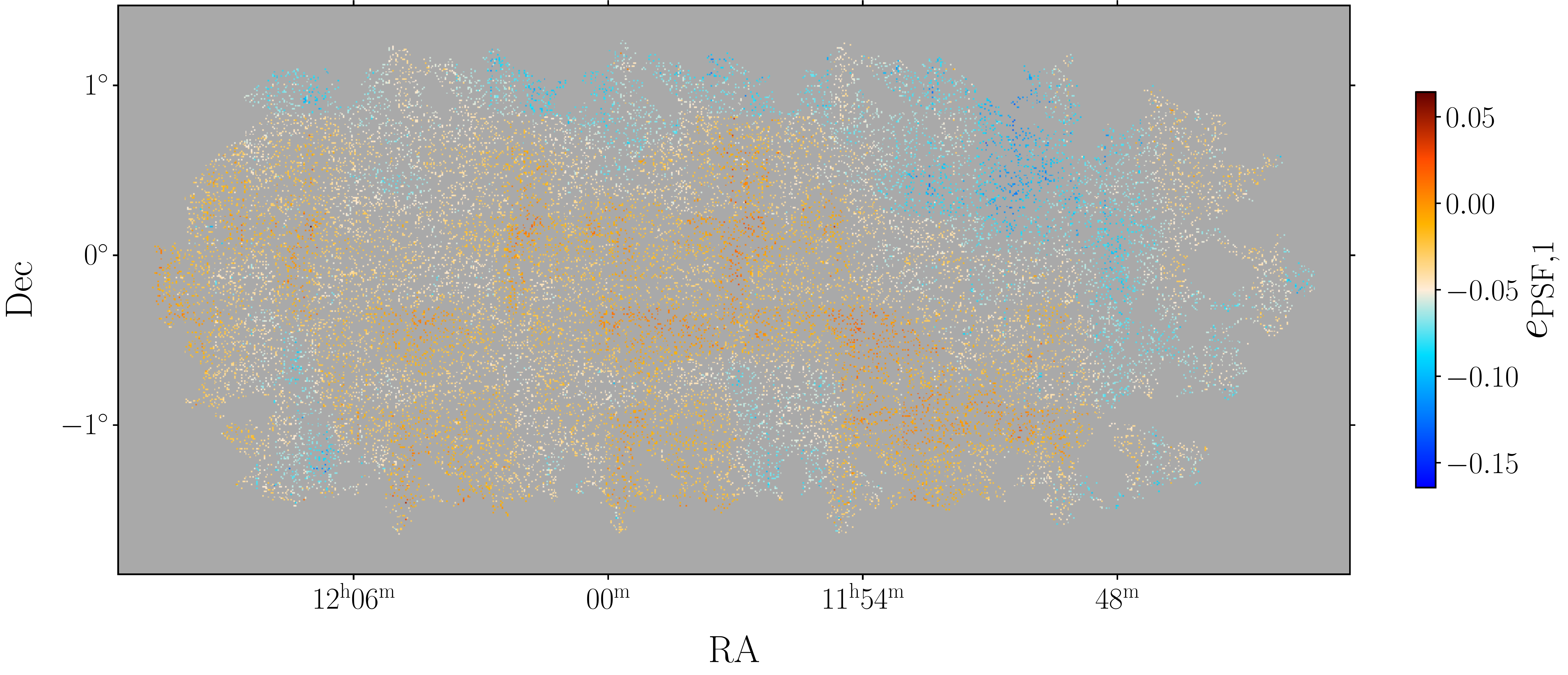}
        \includegraphics[width=0.8\textwidth]{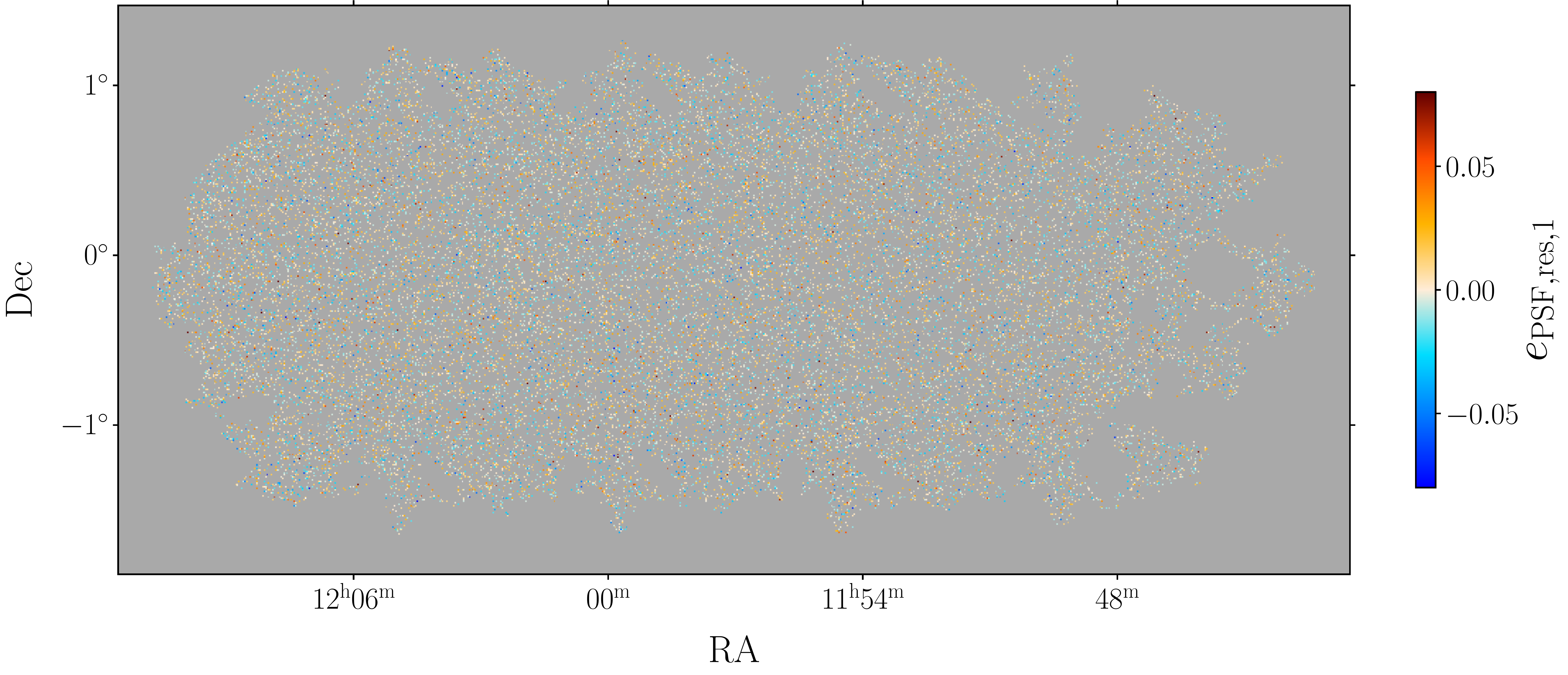}
        \caption{Maps of the galaxy ellipticity, lensing weights, PSF ellipticity and PSF ellipticity residual for the HSC \texttt{WIDE12H} field, which covers approximately 10.5 deg$^{2}$ on the sky. For all maps, we show results for the lowest redshift bin and we only show the first ellipticity component $\ellip_{1}$, where appropriate.
        } \label{fig:hsc.maps}
      \end{figure}
      \begin{figure}
        \centering
        \includegraphics[width=0.9\textwidth]{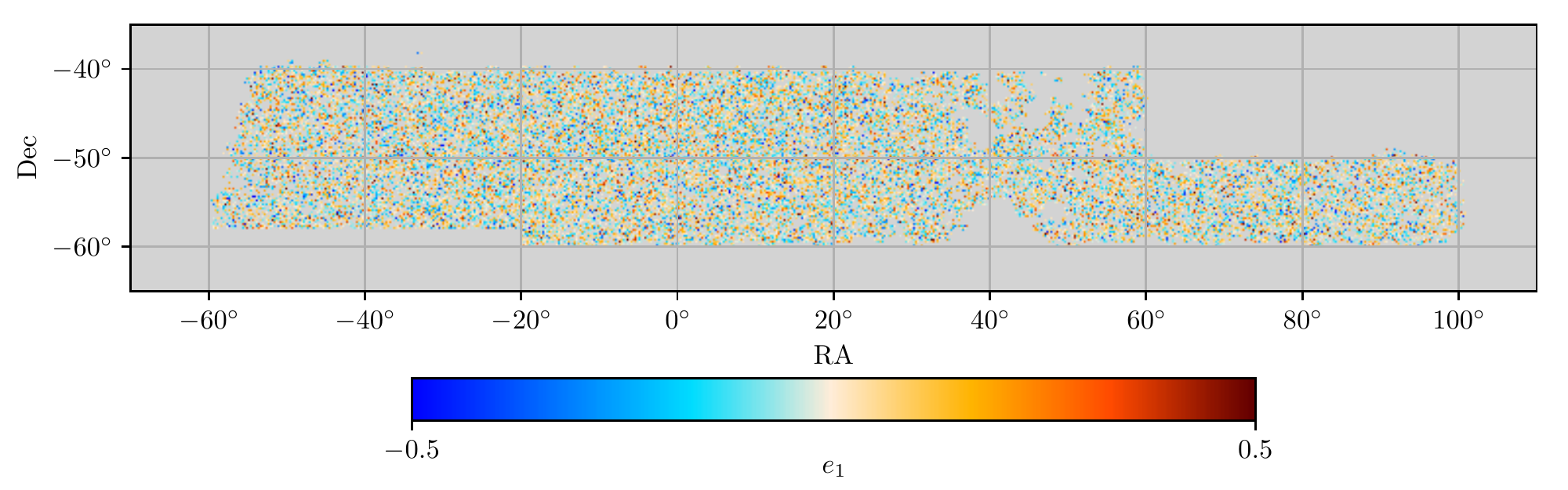}
        \includegraphics[width=0.9\textwidth]{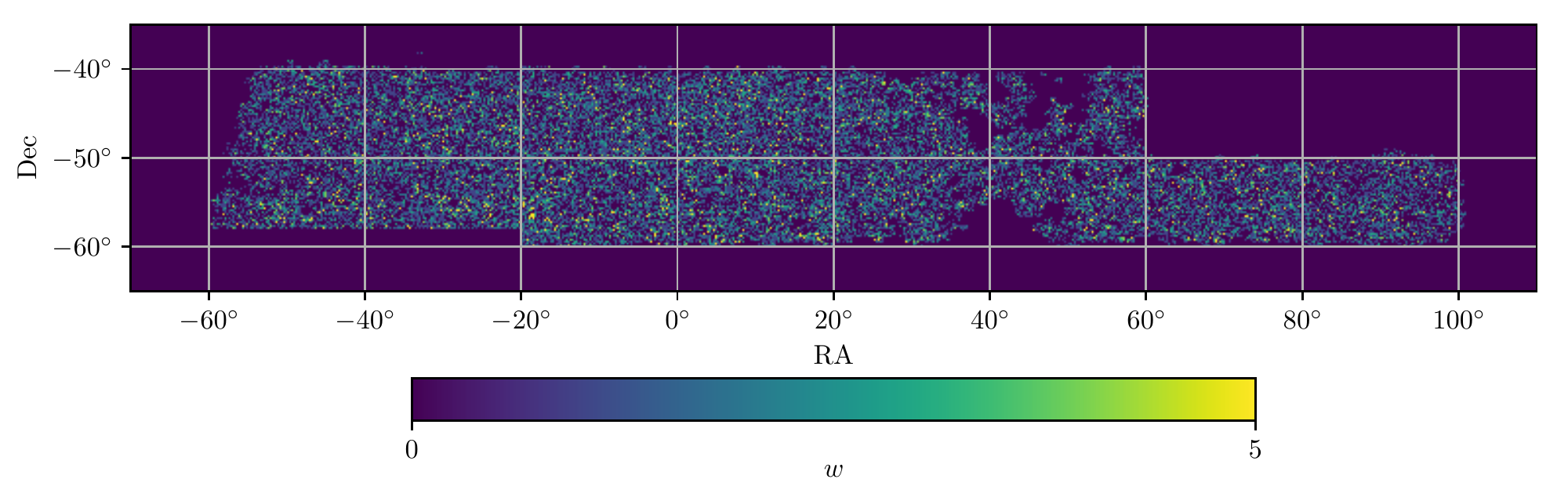}
        \includegraphics[width=0.9\textwidth]{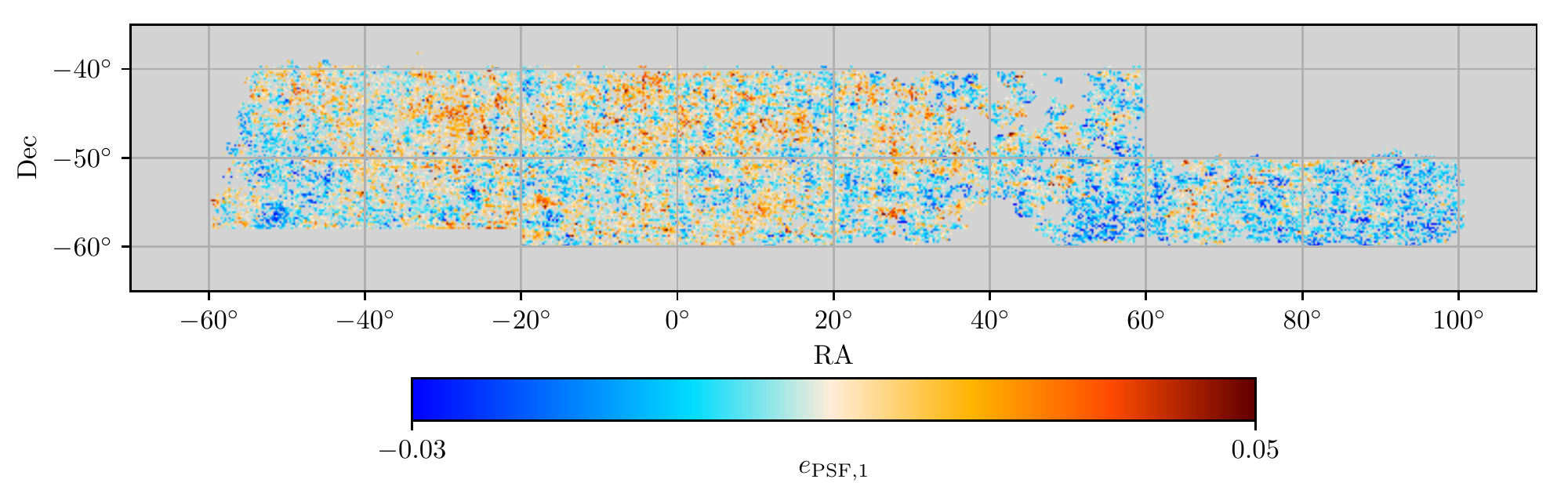}
        \caption{Maps of the galaxy ellipticity, angular mask and PSF ellipticity for DES Y1 data and the second redshift bin. For all ellipticity maps, we only show the first component $\ellip_{1}$. For plotting purposes, the color bar limits are set to the $3\sigma$ values of the field distribution.} \label{fig:des.maps}
      \end{figure}

    \subsection{Measurement pipeline}\label{ssec:results.pipe}
      The first step in estimating shear power spectra is the generation of maps from the corrected galaxy ellipticities (see Eq.~\ref{eq:shear_map}), and their associated weights. In this work, we compute all weight maps using the SW prescription given in Eq.~\ref{eq:weight_map_SW}.

      For HSC, we generate separate shear maps for 5 DR1 fields, namely \texttt{GAMA09H, GAMA15H, WIDE12H, VVDS} and \texttt{XMM}. Given the small area covered by each field, we use the flat-sky approximation to take advantage of efficient fast Fourier transform (FFT) methods. As in Ref.~\cite{1912.08209}, we use a Plate Carr\'ee projection to minimize the curved-sky distortions in the azimuth direction. We use a square pixelization grid with a pixel size of $0.5$ arcminutes. The rectangular map covering each field has a buffer of 24 masked pixels along the edges to minimize boundary effects in the FFTs. 
      \begin{table}
        \centering
        \begin{tabular}{|cccc||cc|}
          \hline
          \multicolumn{4}{|c||}{DES} & \multicolumn{2}{c|}{HSC} \\
          \hline
          0 & 309 & 1098 & 3914 & 100 & 3000\\
          30 & 351 & 1247 & 4444 & 200 & 3800 \\
          60 & 398 & 1416 & 5047 & 300 & 4600\\
          90 & 452 & 1608 & 5731 & 400 & 6200\\
          120 & 513 & 1826 & 6508 & 600 & 7800\\
          150 & 582 & 2073 & 7390 & 800 & 9400\\
          180 & 661 & 2354 & 8392 & 1000 & 12600\\
          210 & 750 & 2673 & 9529 & 1400 & 15800\\
          240 & 852 & 3035 & 10821 & 1800 & \\
          272 & 967 & 3446 & 12288 & 2200 & \\
          \hline
        \end{tabular}
        \caption{Bandpower edges used for the DES (left) and HSC (right) power spectra. Note that only the bandpowers with $\ell<2\Nside=8{,}192$ were used in the case of DES.}\label{tab:bpws}
      \end{table}

      In order to test for systematics, we complement the shear maps with two additional maps: a map of the PSF ellipticity as a function of angular position and a map of the PSF residuals. In HSC, $20\%$ of the star sample is not used for PSF estimation but is set aside for validation purposes. We use this reserved subsample to create  maps of the PSF ellipticity and PSF residuals. Specifically, we estimate PSF residuals as the difference between PSF model and measured ellipticities for stars in the reserved sample. This approach follows Refs.~\cite{1705.06745, 1809.09148}, but we note that we use a star catalog different from the one used in Refs.~\cite{1705.06745, 1809.09148}, as the latter was found to be contaminated by galaxies (H. Miyatake, S. More, private communication).

      For the DES data, we generate full sky maps using the \texttt{HEALPix}\footnote{\url{http://healpix.sourceforge.net}.}~\cite{astro-ph/0409513} pixelization with resolution parameter $\Nside = 4096$, which corresponds to a pixel size of $\sim0.86$ arcmin. The shape measurement weights $v_i$ are by definition $1$ for all sources in the \mcal{} catalog, and thus the weight maps simply trace the number of sources per pixel. We additionally generate maps of the PSF ellipticity to test for systematics in the shear power spectra, following the methods outlined above with the exception of the ellipticity correction step in Eq.~\ref{eq:des:ecal}.

      Figures \ref{fig:hsc.maps} and \ref{fig:des.maps} show the shear maps, weights map and PSF ellipticity maps for both HSC and DES, respectively. For DES, we show the maps for the second redshift bin, while for HSC, we show the maps for the lowest redshift bin and the \texttt{WIDE12H} field. In addition, we also show the map of PSF residuals for HSC.

      From these maps, we estimate power spectra and their associated covariance matrices using the publicly-available code \nmt{}\footnote{\url{https://namaster.readthedocs.io/en/latest}.} \cite{1809.09603}. We use the flat- and curved-sky versions of the code for the HSC and DES data, respectively, and compute all auto- and cross-correlations between different pairs of maps, including different tomographic bins and $E/B$ spin-2 components. The power spectra are computed on a set of consecutive bandpowers with edges listed in Tab.~\ref{eq:bandpowers}: the HSC bandpowers follow a piecewise-linear spacing as in Ref.~\cite{1912.08209}, while the DES bandpowers are linearly spaced up to $\ell=240$ and then follow a constant logarithmic spacing. We use the weight map directly as the mask of each \nmt{} field without any additional apodization, and we do not perform $E$ or $B$-mode purification. The shear power spectrum is not steep enough to benefit from mask apodization and, given the patchiness of the shear weights map, the loss of area incurred by any apodization has a strong detrimental effect on power spectrum uncertainties. Furthermore, since in this case the most interesting signal is the dominant $E$-mode, no benefit can be gained from $E/B$ purification. For HSC, we estimate these power spectra separately for each field and coadd the resulting power spectra, weighted by the sum of the shear weights for each field, to obtain a single spectrum (cf. Ref.~\cite{1809.09148}). In order to estimate the noise bias, we follow the analytical prescription given in Sec.~\ref{ssec:pcl.shear} and we subtract the obtained power spectra from each auto-correlation. We test the stability of the computed power spectra with respect to choice of pixelization by repeating our analysis for different pixel resolutions: for HSC, we increase the pixel size by a factor of two, and for DES, we recompute all power spectra at resolutions of $N_{\mathrm{side}}=512,\,1024,\,2048$, finding consistent results for the angular scales considered in this work in all cases. Finally, we analytically estimate the covariance matrix following the method described in Sec.~\ref{ssec:pcl.covar}, fully accounting for the different mode-coupling coefficients for the signal and noise contributions as in Eq.~\ref{eq:covar_pcl}. The signal power spectra used as input for the covariance calculation for HSC and DES are computed for the cosmological parameters reported by the \planck{} collaboration (Eq. \ref{eq:planckcosmo}).
 
    \subsection{Methods validation}\label{ssec:results.method}
      Before presenting the cosmic shear power spectra measured from the HSC and DES datasets, we validate the methods outlined in Sections \ref{ssec:pcl.noise} and \ref{ssec:pcl.covar} to analytically estimate the noise bias and disconnected covariance matrices. Doing so for HSC and DES will allow us to validate the expressions for both flat-sky and  curved-sky.

      \subsubsection{Noise bias}\label{ssec:results.method.noise}
        \begin{figure}
          \centering
          \includegraphics[width=0.9\textwidth]{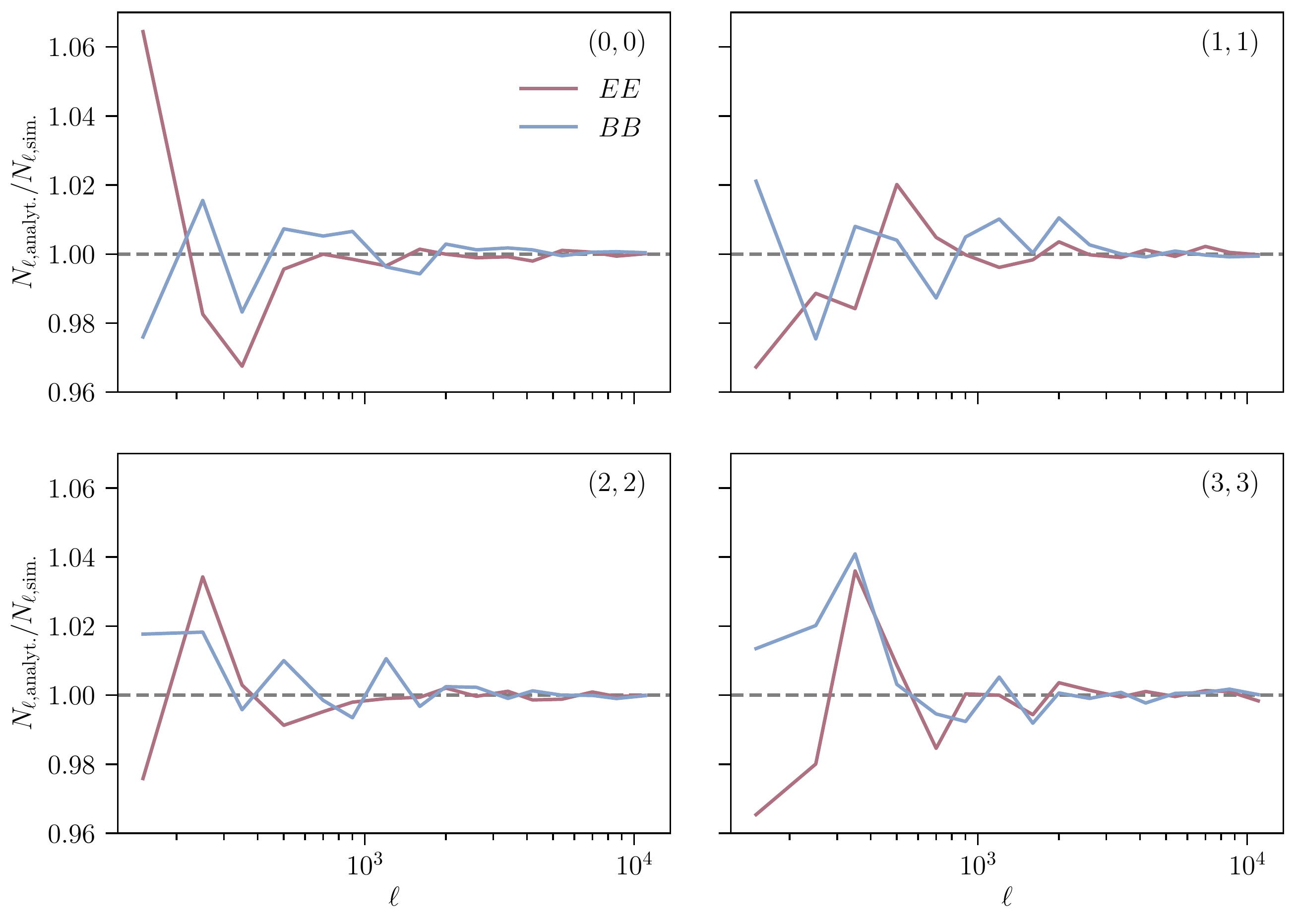}
          \includegraphics[width=0.9\textwidth]{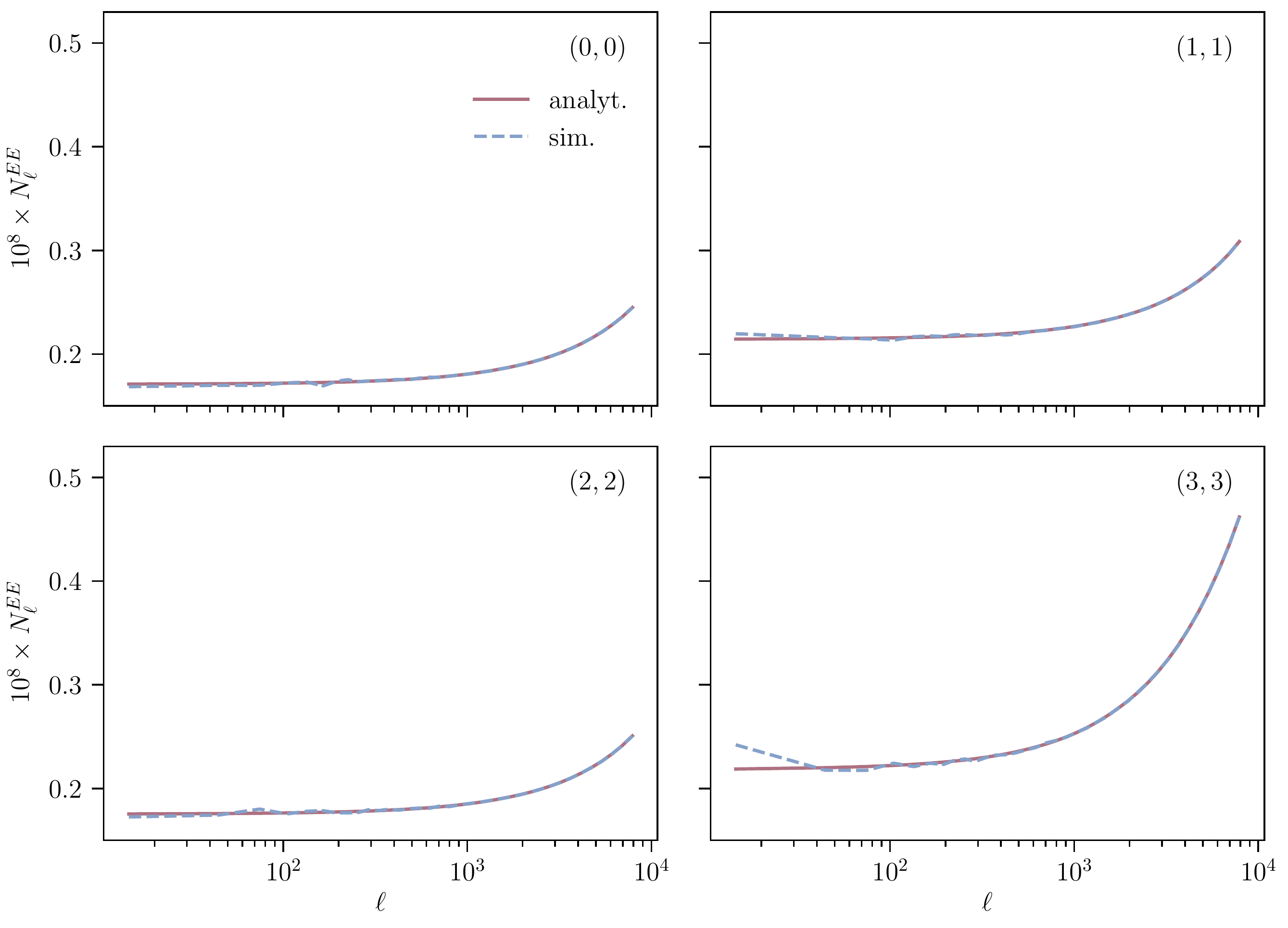}
          \caption{Comparison of the analytic shape noise estimate using Eq.~\ref{eq:noise_bias_shear} with the noise power spectrum obtained from 1000 randomized ellipticity maps for the HSC \texttt{WIDE12H} field (top panel) and DES (bottom panel). Results are shown for the four different redshift bins used in each experiment. For HSC, we show results for the ratio of the analytic and simulated noise bias for both $E$ and $B$-modes, while for DES we only show results for the $E$-mode noise bias (the result for $B$-modes is equivalent).}
          \label{fig:noise_bias}
        \end{figure}
        We validate the analytic calculation of the noise bias described in Sec.~\ref{ssec:pcl.noise} by generating 1000 random realizations of the shear maps for both the HSC and DES samples. For each realization, we create a randomized map by rotating all galaxy ellipticities by a random angle while keeping the galaxy positions fixed. This approach removes ellipticity correlations induced by cosmic shear, while retaining noise fluctuations due to galaxy number density anisotropy and shape noise. The mean of an ensemble of these maps is therefore expected to give a reasonable estimate of the noise power spectrum. In Fig.~\ref{fig:noise_bias}, we compare the mean noise obtained from these realizations to the analytical prediction given in Eq. \ref{eq:noise_bias_shear} for both HSC and DES for all four redshift bins used in the respective analyses. The upper panel of this figure shows the ratio of the analytical noise bias to that estimated from the simulations for the HSC \texttt{WIDE12H} field, which has the smallest sky coverage and is therefore expected to be most affected by mode-coupling and masking uncertainties. The lower panel shows the two estimates of the noise bias for the DES data. As can be seen, we find good agreement between both estimates in both cases, with maximal differences of approximately $5\%$ seen for the largest multipoles. These deviations are due to statistical noise from the finite number of random rotations, rather than inaccuracies in the analytic method, as Eq.~\ref{eq:noise_bias_shear} does not involve any approximations. The expression given in Eq.~\ref{eq:noise_bias_shear} therefore provides a noise estimate at significantly lower computational cost, avoiding the generation of potentially expensive noise realizations\footnote{For reference, computing the spherical harmonic transform of a single spin-2 field with \texttt{HEALPix} resolution parameter $N_{\rm side}=4096$ takes about 2 minutes on an Intel i7 quad-core laptop. Thus computing the noise level from $1{,}000$ random rotations takes about 33 hours on the same system for a single power spectrum, compared with a simple map-averaging operation taking fractions of a second.}. As noted in Section \ref{ssec:pcl.noise}, the scale dependence of the noise power spectrum that can be observed in the lower panel is induced by the inverse mode-coupling matrix and the fact that we are probing a finite range of multipoles $\ell$.

      \subsubsection{Covariance matrices}\label{ssec:results.method.covar}
        To validate the analytical covariance described in Sec.~\ref{ssec:pcl.covar}, we use a suite of $10{,}000$ Gaussian simulations\footnote{In order to be able to generate this many simulations, the curved-sky test was run at a downgraded resolution of $N_{\rm side}=1024$.}. Each realization is comprised of a signal and noise map. We compute the noise map from the data, as described above, and add a Gaussian cosmic shear power spectrum realization, calculated with the Core Cosmology Library ({\tt CCL}, \cite{1812.05995}) for a fiducial set of cosmological parameters, as reported by the \planck{} collaboration\footnote{See the fourth column in Tab. 2 in Ref.~\cite{1807.06209}.}:
        \begin{equation}\label{eq:planckcosmo}
          \cospar{} = (0.3133,\allowbreak 0.0493,\allowbreak 0.6736,\allowbreak 0.9649,\allowbreak 0.8111),
        \end{equation}
        and we further assume vanishing nuisance parameters. In both cases, we additionally employ the redshift distributions described in Section \ref{sec:data} to compute theoretical predictions. We then estimate the covariance from the sample variance of these realizations and compare it to two analytical estimates: first, the standard NKA approximation from Ref.~\cite{1906.11765} (Eq. \ref{eq:nka}), using the unmodified signal power spectrum and the effective noise bias for covariances (Eq. \ref{eq:noicov}). Secondly, the improved NKA approximation (Eq. \ref{eq:nka_plus}), fully accounting for the noise terms as in Eq. \ref{eq:covar_pcl}. Note that we do not consider the case of the standard NKA using the mode-decoupled noise bias instead of the effective noise bias, as we find that this performs poorly in the cases considered in this work, overpredicting the uncertainties by a factor of $\mathcal{O}(1)$.

        \begin{figure}
          \centering
          \includegraphics[width=\textwidth]{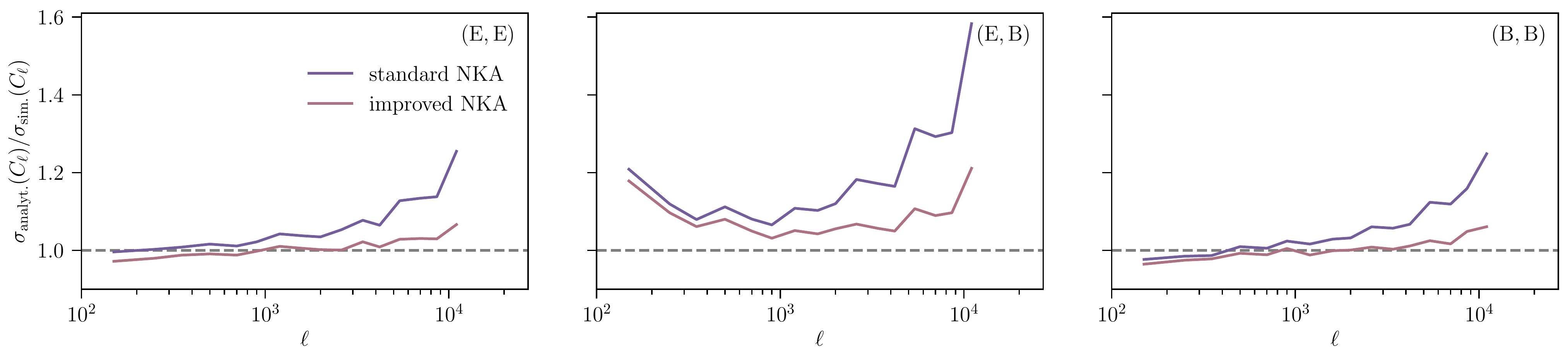}
          \caption{Ratio of the diagonal errors estimated via the standard (Eq.~\ref{eq:nka}) and improved (Eq.~\ref{eq:nka_plus}) NKA to the sample variance errors from a suite of $10{,}000$ Gaussian simulations for the lowest redshift bin in the HSC \texttt{WIDE12H} field.}
          \label{fig:hsc.cov_diag}
        \end{figure}
        \begin{figure}
          \centering
          \includegraphics[width=\textwidth]{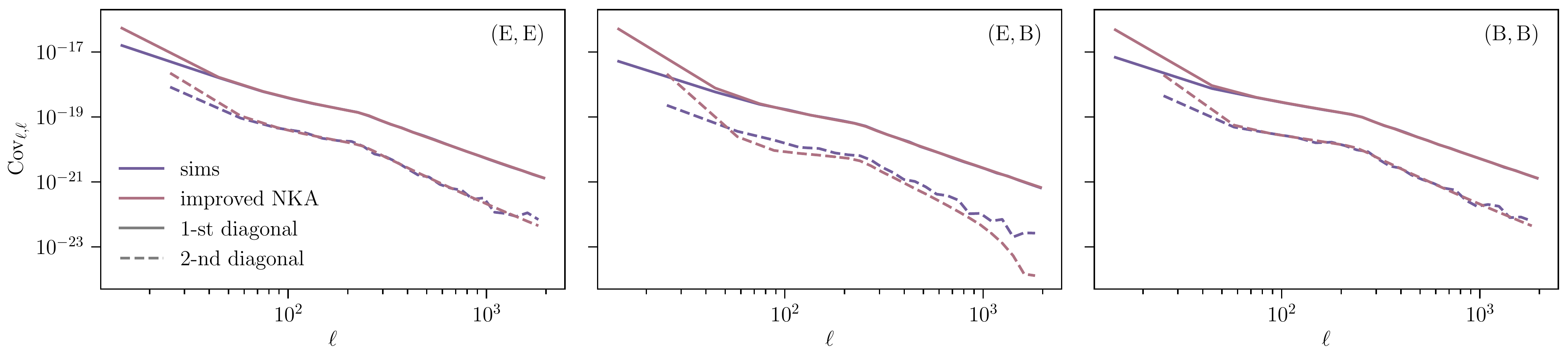}
          \caption{Comparison of the main and second diagonals of the covariance matrix estimated from a suite of Gaussian simulations to both the original analytic estimate presented in Eq.~\ref{eq:nka} and the improved estimate from Eq.~\ref{eq:nka_plus} for the lowest redshift bin in DES.}
          \label{fig:des.cov_diag}
        \end{figure}
        \begin{figure}
          \centering
         \includegraphics[width=\textwidth]{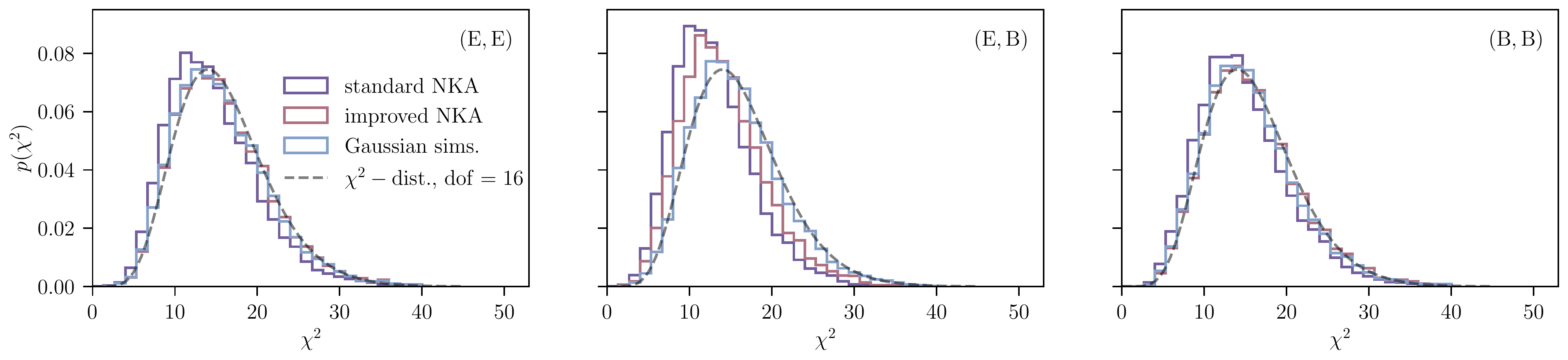}
          \includegraphics[width=\textwidth]{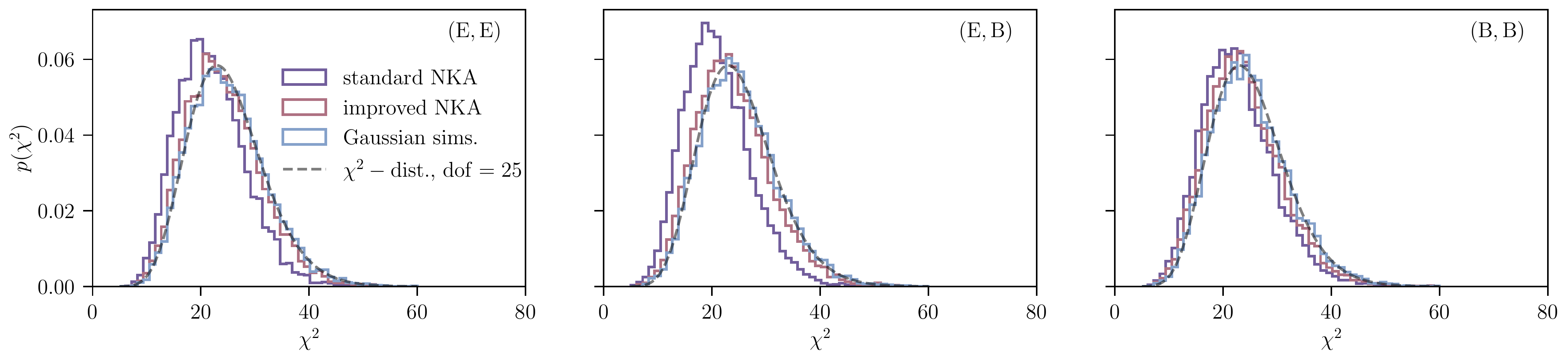}
          \caption{Distributions of $\chi^2$-values obtained using the simulated covariance and the standard and improved NKA (Equations \ref{eq:nka} and \ref{eq:nka_plus} respectively) for the auto-correlation of the lowest redshift bin in the HSC \texttt{WIDE12H} field (upper panel) and the DES Y1 data (lower panel). Results are shown for the $EE$, $EB$ and $BB$ correlations in the left, middle and right panels, respectively.}
          \label{fig:chi2ee_eb_bb}
        \end{figure}
        \begin{figure}
            \centering
            \includegraphics[width=0.65\textwidth]{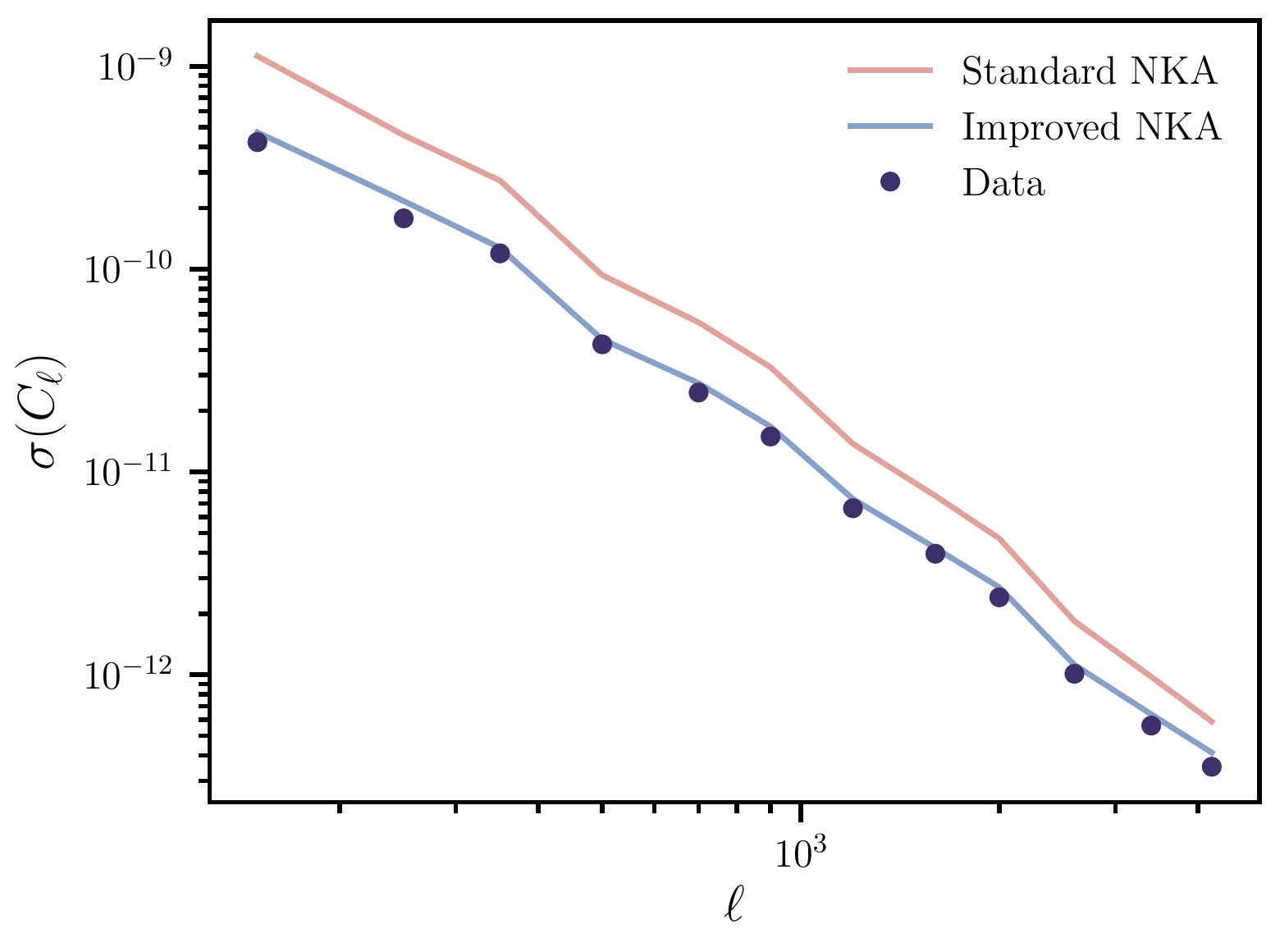}
            \caption{Power spectrum uncertainties obtained from a set of 1000 noise-less Gaussian simulations using the weights map corresponding to the HSC \texttt{WIDE12H} field (dark blue circles). The rose and blue lines show the analytical prediction using the standard narrow-kernel approximation (NKA) and the improved implementation presented in Eq.~\ref{eq:nka_plus}. Since the simulations contain no shape noise, the covariance matrix receives only the first contribution in Eq.~\ref{eq:covar_pcl}, for which we use the improved NKA. Since all the noise terms in that equation are computed exactly, this setup is meant to highlight the inaccuracies of the NKA. While the standard NKA grossly overestimates the signal part of the covariance, the improved implementation recovers the simulated power spectrum uncertainties up to $\sim10\%$ in the range of scales used for cosmological analyses.}\label{fig:covar_noiseless}
        \end{figure}
        The results for the lowest redshift bin in HSC and DES are shown in Figures \ref{fig:hsc.cov_diag} and \ref{fig:des.cov_diag}, respectively. Fig.~\ref{fig:des.cov_diag} shows the first and second diagonals of the $EE$, $EB$ and $BB$ covariances for DES. The use of the improved NKA with the mode-coupled signal power spectrum improves the agreement with the simulated covariance on large, signal-dominated scales. Nevertheless, we find that the analytical uncertainties are always overestimated for the lowest bandpower (in agreement with the results of Ref.~\cite{1906.11765}). As we show later (and in Ref.~\cite{1906.11765}), this has a negligible impact on both goodness-of-fit tests and final parameter constraints. On small scales, the use of the effective noise bias (Eq.~\ref{eq:noicov}) leads to a $\sim10\%$ over estimate of the first diagonal and a severe under-estimation of the second diagonal. This $\sim10\%$ offset is shown explicitly in Fig.~\ref{fig:hsc.cov_diag}, which presents the ratio of the first diagonals of the standard and improved NKA estimators to those of the simulated covariance for the HSC \texttt{WIDE12H} field. For HSC, we find results consistent with those for DES except that we do not see evidence for error overestimation on large angular scales. For both datasets, we find the largest differences between analytic and simulated covariance for the case of the $EB$ power spectrum. This is due to the specific form of the NKA for spin-2 fields, which neglects the spatial derivatives of the weight map \citep{1906.11765}. As shown below, the accuracy of the analytic Gaussian covariance is nevertheless sufficient for the analysis presented in this work, even to derive accurate $\chi^2$-values when using $EB$ correlations as null tests. However, these conclusions might change for surveys with larger angular sky coverage that aim to probe cosmic shear at the largest scales. As the inaccuracies shown in Figures \ref{fig:hsc.cov_diag} and \ref{fig:des.cov_diag} mainly affect the lowest multipoles, these problems could potentially be overcome by using fast, low-resolution Gaussian simulations to estimate covariances at large angular scales, while using analytic estimates in the small-scale regime.
        
        Although the practical effects of the quoted $\sim10\%$ mis-estimation of the power spectrum uncertainties when using the effective noise bias may often be negligible, we recommend that the different mode-coupling coefficients for signal and noise be taken into account in order to obtain unbiased cosmic shear covariances, in particular when considering large angular multipoles, if the computational resources are available.

        We additionally perform a more quantitative comparison between the different covariance matrix approximations described in this work and compute the $\chi^2 \equiv \sum_{\ell\ell'}(C_\ell^{i} - \bar{C}_\ell) {\rm Cov}^{-1}_{\ell\ell'} (C_{\ell'}^{i} - \hat{C}_{\ell'})$ of each simulation (labeled by $i$) with respect to the mean power spectrum of all simulations using the different covariance matrix estimates ($\bar{C}_\ell)$. The distributions of the obtained $\chi^2$-values are shown in Fig.~\ref{fig:chi2ee_eb_bb} for HSC and DES. In all cases, we find that the simulated $\chi^2$-values follow a $\chi^2$ distribution,  as expected. The distributions for the standard NKA estimator are generally offset by $\sim10-15\%$ from the simulations, especially in the curved-sky DES case. The improved NKA estimator reduces this offset significantly to $\lesssim4\%$. As expected, the performance in the case of the $EB$ covariance is worse in both cases, but sufficiently accurate to use $EB$ correlations as a null test for systematics, as we show in Section \ref{ssec:results.val}.
        
        The results presented above involve simulations that contain both the cosmological shear signal and shape noise. Since the noise contributions to the covariance matrix are calculated exactly in Eq.~\ref{eq:covar_pcl}, the inaccuracies of the standard and improved NKA are not readily evident in this setup. To highlight these, we repeat our analysis for 1{,}000 signal-only simulations of the HSC \texttt{WIDE12H} field. As can be seen from Fig.~\ref{fig:covar_noiseless}, we find that the standard NKA over-estimates the signal contribution to the covariance by more than a factor of 2. The improved NKA on the other hand recovers the power spectrum uncertainties obtained from the simulations to within $\sim10\%$ in the range of scales used for cosmological analyses. This accuracy is further improved when including the noise contributions.

        These results show that the improved analytical NKA estimator presented in this work is sufficiently accurate for current cosmological weak lensing analyses. Its potential inaccuracies, as we will show in Section \ref{ssec:results.val}, are negligible for practical applications in parameter inference and null tests. This might not be the case for upcoming Stage IV weak lensing surveys as their statistical power requires knowledge of the covariance matrix to $1-4 \%$ in order to avoid systematic errors in the likelihood. We defer an investigation of this issue to future work.

    \subsection{Null tests and validation}\label{ssec:results.val}
      \begin{table}
      \centering
      \begin{tabular}{|l|cccccccc|}
        \multicolumn{9}{c}{HSC}\\
        \hline
        \zbin & $E_0$ & $E_1$ & $E_2$ & $E_3$ & $B_0$ & $B_1$ & $B_2$ & $B_3$\\
        \hline
        \multirow{2}{*}{$B_0$} & 0.69   & 0.75   & 2.43   & 1.56   & 1.41   & -- & -- & -- \\
                               & 68.1\% & 62.5\% & 1.7\% & 14.1\% & 19.4\% & -- & -- & -- \\
        \hline
        \multirow{2}{*}{$B_1$} & 2.96   & 0.78   & 1.15   & 1.00   & 1.04   & 0.60   & -- & -- \\
                               & 0.4\%  & 60.3\% & 32.6\% & 42.6\%  & 40.1\% & 75.5\% & -- & -- \\
        \hline
        \multirow{2}{*}{$B_2$} & 1.63   & 0.83   & 1.23   & 0.69   & 0.92   & 1.73   & 0.36   & -- \\
                               & 12.2\% & 56.2\% & 28.3\% & 68.0\% & 49.1\% & 9.7\% & 92.5\% & -- \\
        \hline
        \multirow{2}{*}{$B_3$} & 1.09   & 2.15   & 1.11   & 0.13   & 1.46   & 0.90  & 0.73   & 2.28   \\
                               & 36.6\% & 3.5\% & 35.4\% & 99.6\% & 17.8\% & 50.6\% & 64.8\% & 2.5\% \\
        \hline
        \multicolumn{9}{c}{}\\
        \multicolumn{9}{c}{DES}\\
        \hline
        \zbin & $E_0$ & $E_1$ & $E_2$ & $E_3$ & $B_0$ & $B_1$ & $B_2$ & $B_3$\\
        \hline
        \multirow{2}{*}{$B_0$} & 1.13   & 0.98   & 0.82   & 0.97   & 0.99   & -- & -- & -- \\
                               & 27.5\% & 51.8\% & 77.0\% & 51.7\% & 48.6\% & -- & -- & -- \\
        \hline
        \multirow{2}{*}{$B_1$} & 1.80   & 0.90   & 1.05   & 1.03   & 0.95   & 0.94   & -- & -- \\
                               & 0.3\%  & 64.0\% & 39.5\% & 41.3\%   & 56.4\% & 58.1\% & -- & -- \\
        \hline
        \multirow{2}{*}{$B_2$} & 1.12   & 0.57   & 0.68   & 1.20   & 0.97   & 0.57   & 1.04   & -- \\
                               & 29.3\% & 98.1\% & 92.8\% & 19.1\% & 52.6\% & 98.3\% & 40.4\% & -- \\
        \hline
        \multirow{2}{*}{$B_3$} & 0.83   & 0.85   & 0.82   & 0.73   & 0.92   & 1.53  & 1.04   & 0.63   \\
                               & 75.9\% & 72.8\% & 76.6\% & 88.7\% & 60.6\% & 2.2\% & 40.6\% & 95.7\% \\
        \hline  
      \end{tabular}
      \caption{$\sfrac{\chi^2}{\dof}$ and associated $p$-values for all $B$-mode null-tests for HSC (top) and DES (bottom). Each row contains all non-repeated cross-correlations with a $B$-mode map for the four redshift bins considered. In each cell, the upper and lower numbers correspond to the reduced $\chi^2$ and $p$-value as a percentage, respectively. Three (two) of the null tests yield $p$-values below 5\% for HSC (DES), compatible with the look-elsewhere effect. In the case of DES, only scales $\ell < 2\Nside=8{,}192$ are included, while we restrict the multipole range to $300 \leq \ell \leq 2000$ for HSC.}
      \label{tab:eb_null}
      \end{table}

      To validate the computed power spectra and covariance matrices, we perform a number of null and validation tests, as outlined below.
        
      Although a number of effects, such as intrinsic alignments or lensing magnification \cite{astro-ph/0202411,0910.3786,0904.4703}, could give rise to $B$-modes in the shear power spectrum, their amplitude should be subdominant, and negligible given the sensitivity of current imaging datasets. Therefore, correlations involving shear $B$-modes have traditionally been used as a null test to identify potential systematics in the data. We estimate the significance of each individual power spectrum involving a shear $B$-mode field by computing its $\chi^2$ with respect to zero and its associated $p$-value. For four tomographic bins there are a total of 26 such null tests, including both $BB$ and $EB$ correlations. The resulting $\chi^2$ and $p$-values are listed in Tab.~\ref{tab:eb_null} for both HSC and DES and the corresponding null spectra are shown in Appendix \ref{app:nulls}. We find only 3 null tests in HSC and 2 in DES with $p$-values below $0.05$. This is not unexpected given the large number of null tests performed. To quantify the significance of these low $p$-values in the context of the look-elsewhere effect, we perform a Kolmogorov-Smirnoff (KS) test of the recovered $\chi^2$-values against a $\chi^2$ distribution with the number of degrees of freedom given by the number of bandpowers in both datasets (7 for HSC, corresponding to bandpowers with $300 \leq \ell \leq 2{,}000$, and 36 for DES with $\ell < 2\Nside = 8{,}192$). Both KS tests pass, with $p$-values of $20\%$ and $54\%$ for HSC and DES, respectively. We verify that the small over-estimation of the covariance matrix for the $EB$ power spectra noted in Section \ref{ssec:results.method.covar} does not affect these conclusions. To do so, we artificially inflate all $\chi^2$-values for the $EB$ null tests by $10\%$ and recompute the associated $p$-values as well as the KS test, finding consistent results. Finally, the reduced $\chi^2$ and associated $p$-value for a null data vector constructed by stacking all individual $BB$ and $EB$ correlations is $\sfrac{\chi^2}{{\rm d.o.f.}}=\sfrac{224.2}{182}=1.23\;(p=1.9\%)$ and $\sfrac{\chi^2}{{\rm d.o.f.}}=\sfrac{923.8}{936}=0.99\;(p=60.1\%)$ for HSC and DES, respectively. Although the $p$-value for the HSC data is relatively low, we note that it improves in the extended scale range $300 \leq \ell \leq 12,000$ to $\sfrac{\chi^2}{{\rm d.o.f.}}=\sfrac{410.1}{364}=1.09\;(p=12.4\%)$. Because of this, and since the distribution of the individual $\chi^{2}$-values is consistent with expectations when using the reduced set of multipoles, we attribute this to a statistical fluctuation. We therefore conclude that there is no evidence for systematics in either dataset giving rise to $B$-modes within the range of scales probed here, in agreement with previous studies by Ref.~\cite{1809.09148} for HSC and Refs.~\cite{1708.01533,1811.10596} for DES.

      Both PSF deconvolution errors as well as PSF modeling errors can lead to biases in inferred galaxy ellipticities. We thus perform an additional test for systematics in the DES data by computing the cross-correlation between the cosmic shear maps and the PSF ellipticity maps. For each redshift bin, we compute the four possible cross-correlations between the $E$ and $B$ components of both maps, and calculate their $\chi^2$-values with respect to zero as well as the associated $p$-values. To estimate the covariance of these cross-correlations, we use the analytic approach described in Sec.~\ref{ssec:pcl.covar}. This requires an estimate of the auto- and cross-correlations for all fields involved. For the PSF ellipticity auto-spectra we use an interpolated version of the spectrum measured from the data, and set all cross-correlations with the cosmic shear field to zero. The resulting $\chi^{2}$-values are given in Tab.~\ref{tab:des.psf}, and the corresponding spectra are shown in Appendix \ref{app:nulls}. As before, the $\chi^2$-values found for these null spectra are reasonable and pass a Kolmogorov-Smirnoff test against a $\chi^2$ distribution with $p=0.73$.

    \begin{table}
      \centering
      \begin{tabular}{|lcccc|}
        \hline
        \zbin & $EE$ & $EB$ & $BE$ & $BB$\\
        \hline
        \multirow{2}{*}{0} & 0.88 & 1.04 & 0.86 & 0.85\\
                           & 67.5\% & 41.0\% & 70.0\% & 72.6\% \\
        \hline
        \multirow{2}{*}{1} & 1.08 & 0.99 & 0.55 & 0.67\\
                           & 33.6\% & 48.1\% & 98.8\% & 93.8\% \\
        \hline
        \multirow{2}{*}{2} & 1.06 & 1.49 & 0.69 & 1.12\\
                           & 36.8\% & 2.96\% & 92.1\% & 28.4\% \\
        \hline
        \multirow{2}{*}{3} & 0.90 & 1.27 & 0.56 & 1.06\\
                           & 64.4\% & 13.0\% & 98.4\% & 37.9\% \\
        \hline
      \end{tabular}
      \caption{$\sfrac{\chi^2}{\dof}$ and associated $p$-values for all possible cross-correlations between galaxy and PSF ellipticities for DES Y1 and angular multipole range $\ell < 2\Nside=8{,}192$.}\label{tab:des.psf}
    \end{table}
    
      In order to assess the impact of PSF systematics on the HSC cosmic shear signal measured in this work, we follow Ref.~\cite{1809.09148} and make the following Ansatz for the measured galaxy shears
      \begin{equation}
        \hat{\boldsymbol{\gamma}} = \boldsymbol{\gamma} + \alpha \ellip_{\mathrm{PSF}} + \beta \ellip_{\mathrm{PSF}, \mathrm{res}},
      \end{equation}
      i.e. we assume a linear leakage of PSF ellipticity $\ellip_{\mathrm{PSF}}$ and PSF residuals $\ellip_{\mathrm{PSF},\mathrm{res}}$ into the measured galaxy shears $\hat{\boldsymbol{\gamma}}$. In this model, the observed cosmic shear power spectrum is given by
      \begin{equation}
        C_{\ell}^{\hat{\gamma}\hat{\gamma}} = C_{\ell}^{\gamma\gamma} + \alpha^2 C_{\ell}^{e_{\mathrm{PSF}}e_{\mathrm{PSF}}} + 2 \alpha \beta C_{\ell}^{e_{\mathrm{PSF}}e_{\mathrm{PSF}, \mathrm{res}}} + \beta^2 C_{\ell}^{e_{\mathrm{PSF}, \mathrm{res}}e_{\mathrm{PSF}, \mathrm{res}}}.
      \end{equation}
      We estimate the PSF leakage power spectra using the systematics maps described in Sec.~\ref{ssec:results.pipe}. Specifically, we compute the cross-correlations
      \begin{align}
        C_{\ell}^{\hat{\gamma}e_{\mathrm{PSF}}} &= \alpha C_{\ell}^{e_{\mathrm{PSF}}e_{\mathrm{PSF}}} + \beta C_{\ell}^{e_{\mathrm{PSF}}e_{\mathrm{PSF}, \mathrm{res}}}, \\
        C_{\ell}^{\hat{\gamma}e_{\mathrm{PSF}, \mathrm{res}}} &= \alpha C_{\ell}^{e_{\mathrm{PSF}, \mathrm{res}}e_{\mathrm{PSF}}} + \beta C_{\ell}^{e_{\mathrm{PSF}, \mathrm{res}}e_{\mathrm{PSF}, \mathrm{res}}},
      \end{align}
      for all galaxies (i.e. we do not subdivide the sample into redshift bins) and estimate $\alpha, \; \beta$ using a weighted least squares fit in the angular multipole range used in our analysis\footnote{We note that the fit deteriorates significantly if we include bandpowers at angular multipoles smaller than $\ell_{\mathrm{min}} = 300$. This suggests the presence of PSF leakage into the cosmic shear power spectra at these angular scales and we therefore do not include bandpowers smaller than $\ell_{\mathrm{min}} = 300$ in our analysis. These findings are consistent with Ref.~\cite{1809.09148}.}. We find $\alpha = 0.031 \pm 0.029$, $\beta = -0.71 \pm 1.10$, where the uncertainties have been derived from the least squares fit assuming errors on the power spectra given by the analytical $f_{\mathrm{sky}}$ approximation (see e.g. \cite{9305030}). The resulting noise-removed PSF leakage power spectra are shown in Fig.~\ref{fig:hsc.psf} alongside the non-tomographic cosmic shear power spectrum. As can be seen, we find the PSF leakage to be subdominant compared to the non-tomographic cosmic shear signal for the angular scales considered in this analysis. We therefore conclude that the cosmic shear power spectra are not significantly affected by PSF systematics. However, we note that these systematics might become more important for tomographic  power spectra and should therefore be marginalized over in a cosmological analysis, as was done in Ref.~\cite{1809.09148}. Alternatively, these types of systematics can also be addressed at the map level through template deprojection \citep{1809.09603}.
     
      Finally, we validate the $E$-mode shear power spectra and associated covariances obtained in this work against earlier results by assessing their compatibility with cosmological predictions derived by both collaborations. This comparison is contrary to the blinding process that is now commonplace in standard weak lensing cosmological analyses. The data presented here, however, have already been used by the HSC and DES collaborations to extract cosmological constraints, and therefore this comparison should not invalidate the usefulness of these measurements. Note that we do not make any attempt to derive cosmological constraints from these data, a task that we leave for future work.
        
      For HSC, we compute the $\chi^2$ and associated $p$-value of the $E$-mode power spectra with $300 \leq \ell \leq 2000$ with respect to the theoretical predictions derived using a cosmological model compatible with the HSC results:
      \begin{equation}\label{eq:hsccosmo}
        \cospar{} = (0.162,\allowbreak 0.0335,\allowbreak 0.81,\allowbreak 0.96,\allowbreak 1.056),
      \end{equation}
      fixing all nuisance parameters to zero. 

      For DES, we similarly compute theoretical predictions for a model compatible with DES results:
      \begin{equation}\label{eq:descosmo}
        \cospar{} = (0.26,\allowbreak 0.0479,\allowbreak 0.685,\allowbreak 0.973,\allowbreak 0.821),
      \end{equation}
      again setting all nuisance parameters to zero.\footnote{We note that the best-fit cosmological parameters found by the two collaborations appear significantly different at face value. We attribute these differences to the fact that cosmic shear tightly constrains $S_{8} \coloneqq (\sfrac{\Omega_{m}}{0.3})^{\alpha}$, while being largely insensitive to the remaining standard cosmological parameters. Setting $\alpha = 0.45$ and comparing the values obtained for $S_{8}$ with both models, we find them to agree within the quoted uncertainties. It will however be interesting to test the consistency between HSC and DES by performing a joint analysis of the surveys, which we leave for future work.} For both HSC and DES, we then compute the $\chi^2$ and associated $p$-value of the full $EE$ data vector against the theoretical prediction, including scales $\ell\leq2000$ and using the full non-Gaussian covariance matrix. We note, however, that in both cases we find the impact of the non-Gaussian contributions to the covariance to have a negligible $\lesssim10\%$ effect on the $\chi^2$-values.

      \begin{figure}[htb]
        \centering
        \includegraphics[width=\textwidth]{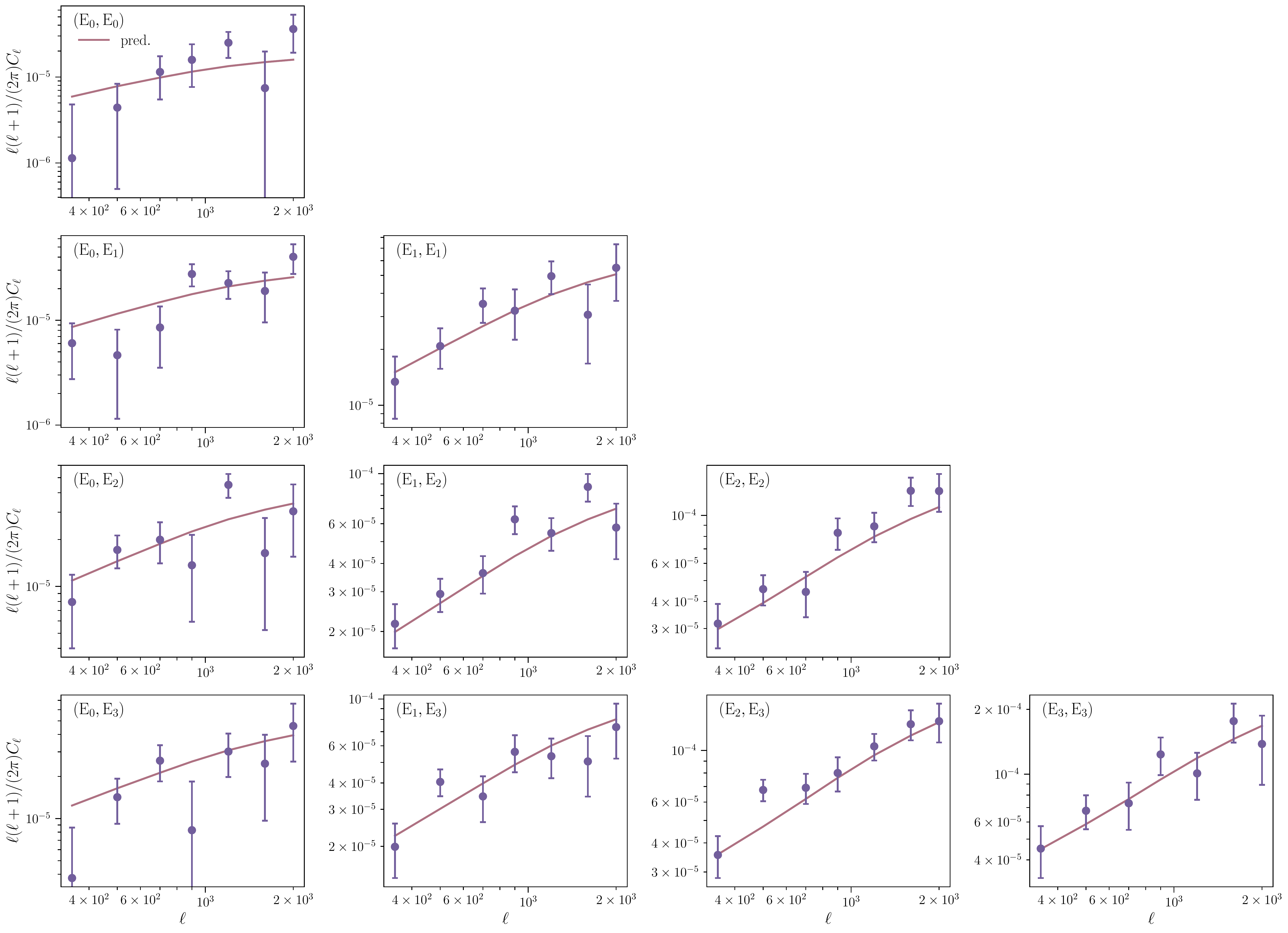}
        \caption{Angular $E$-mode power spectra for HSC. The solid lines show the theoretical predictions derived from the fiducial model given in Eq. \ref{eq:hsccosmo}.}
        \label{fig:hsc.Cls_ee}
      \end{figure}
      \begin{figure}[htb]
        \centering
        \includegraphics[width=\textwidth]{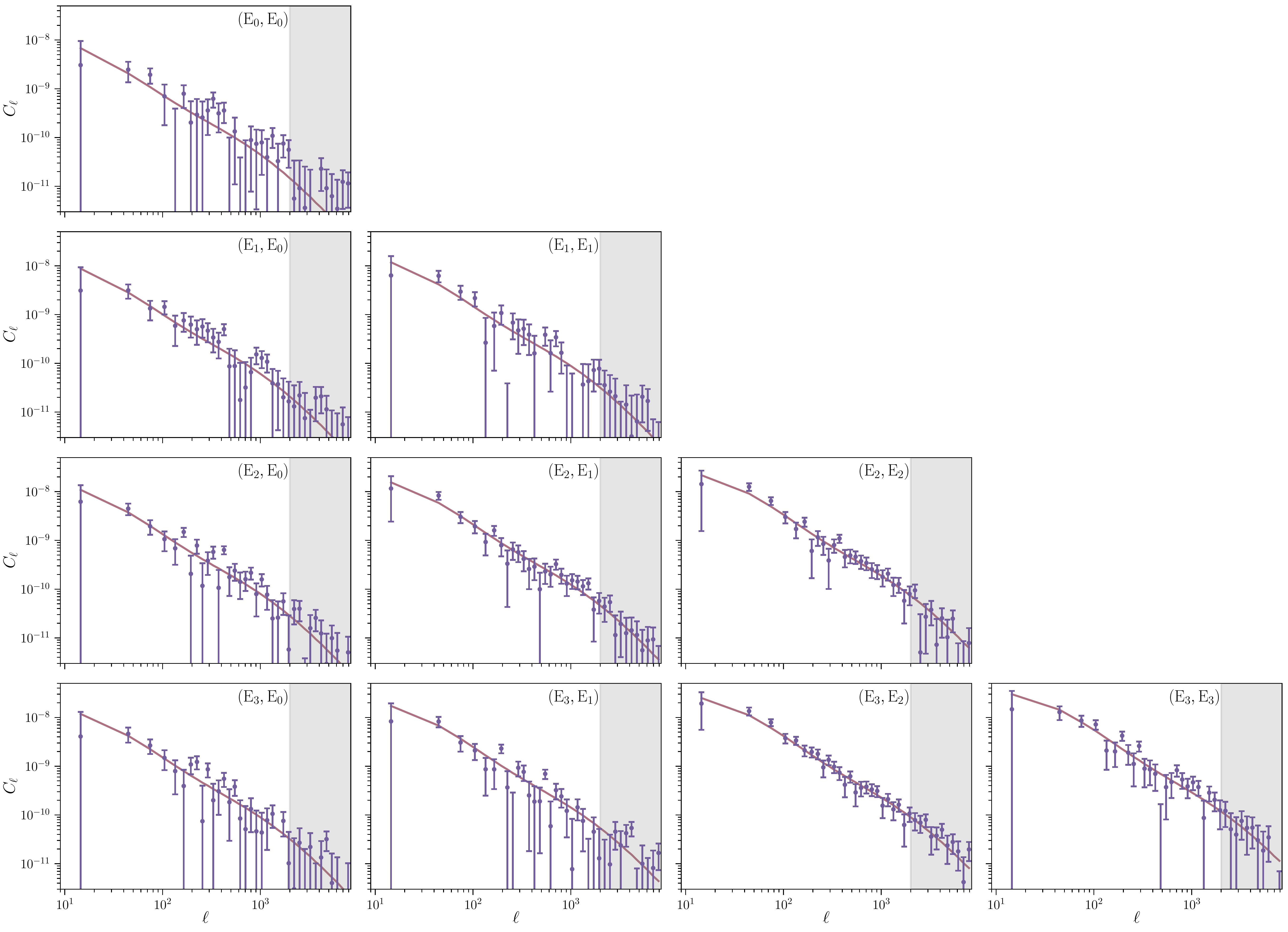}
        \caption{Angular $E$-mode power spectra for DES Y1. The solid lines show the theoretical predictions derived from the fiducial model given in Eq. \ref{eq:descosmo}. The shaded regions indicate angular scales not used in our analysis.}
        \label{fig:des.Cls_ee}
      \end{figure}
      
      The resulting reduced $\chi^{2}$'s for both HSC and DES are
      \begin{align}
        &\sfrac{\chi^2_{\rm HSC}}{\dof}=1.27,\hspace{12pt}(\dof=70,\,\,p=6.4\%),\\
        &\sfrac{\chi^2_{\rm DES}}{\dof}=1.00,\hspace{12pt}(\dof=250,\,\,p=50.6\%).
      \end{align}
      The corresponding $E$-mode power spectra for HSC and DES are shown in Figures \ref{fig:hsc.Cls_ee} and \ref{fig:des.Cls_ee}, alongside the theoretical predictions for the models discussed above. These results are not affected by the over-estimation of the variance of the first bandpower in the curved-sky calculation noted in Section \ref{ssec:results.method.covar}. We verify this by artificially reducing the covariance matrix element for the first bandpower of all auto- and cross-correlations by a factor of $2$ and recomputing the $\chi^2$, which changes only by $\sim0.4\%$. This is in agreement with the results shown in Ref.~\cite{1906.11765}.

      These results therefore show that the measured spectra are in reasonable agreement with the standard cosmological model, and that their uncertainties are not significantly over- or under-estimated. 

   \subsection{Public release}\label{ssec:results.release}
     Having validated the estimated power spectra, we make all measurements presented in this work publicly available through a {\tt github} repository\footnote{ \url{https://github.com/xC-ell/ShearCl}.}. The data are provided as {\tt FITS} files \cite{1981A&AS...44..363W}, and contain all the cosmic shear auto- and cross-correlations (including all $E/B$ combinations), their associated bandpower window functions and covariance matrix, as well as the redshift distributions of the HSC and DES tomographic bins. Therefore, the files contain all the information necessary to carry out a full cosmological analysis of these data. The files were generated using the {\tt SACC} software package \cite{sacc}\footnote{\url{https://github.com/LSSTDESC/sacc}.}. {\tt SACC} is designed to facilitate storage and interpretation of large numbers of two-point correlations in both real and Fourier space, and provides functions to easily select sections of the full data vector and the covariance matrix in a consistent manner. Alongside the data, we further provide a set of {\tt jupyter} notebooks demonstrating the usage of {\tt SACC}, as well as an example analysis pipeline for reproducing the DES shear power spectra from the public catalogs.
        
     For each of the two datasets, we provide two separate files, labeled ${\tt signal\_covG}$ and ${\tt noise\_covNG}$. The ${\tt signal\_covG}$ files contain the measured power spectra, their associated Gaussian covariance, bandpower window functions and redshift distributions. The ${\tt noise\_covNG}$ files contain the estimated noise bias as a data vector and the non-Gaussian contributions to the covariance matrix. The data are stored in the same order in both files so combining them is straightforward. The non-Gaussian contributions to the covariance were calculated using a halo-model approach as described in Ref.~\cite{1601.05779}, following the implementation in Ref.~\cite{2006.00008}, and are provided for convenience.

  \section{Conclusions}\label{sec:conclusion}
    Cosmic shear is one of the most promising tools to constrain the growth of structure and the properties of the late-time accelerated expansion of the Universe. The now standard tomographic analysis of cosmic shear datasets involves the estimation of two-point correlators of galaxy ellipticities in different redshift bins, their forward modeling from theoretical predictions, and constraining the free parameters of the model through a multivariate Gaussian likelihood. Beyond the larger problem of modeling the different systematic uncertainties associated with cosmic shear data (intrinsic alignments, baryonic effects, photometric redshift uncertainties), a careful numerical calculation of all the ingredients of this recipe (unbiased two-point estimators, accurately binned theory predictions, and covariance matrix) is necessary in order to obtain reliable parameter constraints. In this work, we have described in detail how to do so through the use of Fourier-space angular power spectra.

   Fourier-space pipelines offer several advantages over the more traditional real-space analyses, such as robust scale cuts, the avoidance of Hankel transforms, and simpler covariance matrices. However, the calculation of cosmic shear power spectra involves its own challenges: cosmic shear stands out from other projected cosmological probes by the complexity of its sky window function. Since the lensing shear field is only measured through galaxy ellipticities at the respective galaxy positions, the window function is highly inhomogeneous, and essentially given by a sum of Dirac delta functions at those locations. This causes a higher level of statistical coupling between different Fourier modes than is the case in other cosmological observables, which means that some of the aspects of power spectrum estimation must be rigorously handled. In this work, we have focused on two problems: the estimation of the noise power spectrum, or {\sl noise bias}, which must be subtracted from all shear auto-correlations, and the estimation of the {\sl Gaussian covariance matrix}, accounting accurately for all mode-coupling effects. We have shown that both of these quantities can be computed accurately using fast analytical methods that avoid the use of expensive random realizations (e.g. randomly rotated ellipticities, or Gaussian simulations). The main results from this part of the paper are:
    \begin{itemize}
      \item A fast and exact analytical estimate of the noise bias in cosmic shear data that fully accounts for the inherent inhomogeneity of the noise properties of these datasets. This is summarized by Equations \ref{eq:noise_bias} and \ref{eq:noise_bias_shear}. Although this method cannot account for any source of correlated shape noise, neither can the use of randomly rotated catalogs, or the usual avoidance of the zero-lag correlation in real-space analyses. The impact of correlated noise must be either forward-modeled or estimated via other means for both types of analyses. 
      \item Accurate methods to compute the Gaussian covariance matrix. We have shown that the estimate of the signal part of the covariance through the so-called \emph{Narrow Kernel Approximation} can be improved upon significantly by the use of the mode-coupled signal power spectrum in lieu of the true underlying spectrum. This was discussed in Sec.~\ref{sssec:pcl.covar.nkaplus}. Furthermore, we have provided two estimates of the noise contributions to the total covariance. First, Eq.~\ref{eq:covar_pcl} allows for an exact calculation of the noise contribution, which involves the computation of four additional sets of mode-coupling coefficients, involving products of noise variance maps and sky window functions. Secondly, we have shown that the contribution of these additional terms can be accounted for with a reasonable $\sim10\%$ accuracy by adding an effective noise power spectrum to the signal power spectrum in the signal-only contribution, given by Equations \ref{eq:noicov}, \ref{eq:noicov_shear1} and \ref{eq:noicov_shear2}.
    \end{itemize}

    We have then applied these methods to the public data releases from HSC DR1 and DES Y1, which present practical examples of this methodology applied to current data on flat and curved sky, respectively. We have shown that we are able to recover shear power spectra that pass a rigorous set of null tests. In both cases, we observe no evidence of $B$-modes, a common smoking gun for systematic effects, and we determine that any contamination to the power spectra from PSF uncertainties is negligible within the range of scales used here. We find that the distribution of probabilities for all null tests performed is consistent with the expected $\chi^2$ distribution. Furthermore, we find the cosmic shear power spectra obtained in this work to be in good agreement with theoretical predictions derived for the cosmological parameters found by both collaborations. These results indicate that the estimated covariance matrix is not significantly under- or over-estimated. We make these data, together with all additional information (non-Gaussian covariances, redshift distributions, bandpower window functions) needed to conduct a full cosmological analysis, publicly available under \url{https://github.com/xC-ell/ShearCl}. The data are provided in a flexible format especially designed for the storage and interpretation of cosmological two-point function data. The repository also hosts a full analysis pipeline that implements the methods described here on the DES data.

    A number of caveats found in our analysis must be noted. Most notably, the improved NKA used here overestimates the covariance on the first bandpower for the curved-sky calculation. We have found this to have a negligible effect on the data analyzed here, given the small amount of information contained in the lowest multipoles. Fortunately, if an accurate estimate of the large-scale uncertainties is needed, it can be achieved cheaply through the use of low-resolution Gaussian simulations. The approximations involved in the NKA also lead to inaccuracies in the estimates of the $EB$ power spectrum uncertainties, which do not affect the $EE$ and $BB$ errors. These inaccuracies are small, however, and should not preclude the detection of significant systematics in the data. While we find the inaccuracies related to the NKA to be subdominant for current weak lensing data, this might cease to be true for future Stage IV surveys. We defer an investigation of these issues to future work. Finally, care should be taken when using the smallest scales in the estimated power spectra, as both the power spectra and their uncertainties can suffer from edge effects and numerical errors in the Fourier/spherical harmonic transforms. These effects are well isolated, however, and can simply be avoided by discarding the last bandpower in flat-sky analyses, or all multipoles with $\ell>2N_{\rm side}$ if using {\tt HEALPix}.

    Our results show that, in spite of the complications inherent to cosmic shear, the Fourier-space analysis of these data is robust, feasible and practical for both current and future datasets.

  \section*{Acknowledgements}
    We would like to thank Hironao Miyatake and Surhud More for many helpful discussions. Furthermore, we would like to thank Eli Rykoff and the HSC Executive Board for providing access to the star catalog used in this work. C.G.G. is supported by PGC2018-095157-B-I00 from Ministry of Science, Innovation and Universities of Spain and by the Spanish grant, partially funded by the ESF, BES-2016-077038. DA acknowledges support from the Beecroft Trust, and from the Science and Technology Facilities Council through an Ernest Rutherford Fellowship, grant reference ST/P004474. PGF has received funding from the European Research Council (ERC) under the European Union’s Horizon 2020 research and innovation programme (grant agreement No 693024). JD and AN are supported by NSF grant AST-1814971. The Flatiron Institute is supported by the Simons Foundation. This publication arises from research funded by the John Fell Oxford University Press Research Fund. 
    We made extensive use of the {\tt numpy} \citep{oliphant2006guide, van2011numpy}, {\tt scipy} \citep{2020SciPy-NMeth}, {\tt astropy} \citep{astropy:2013, astropy:2018}, {\tt healpy} \citep{Zonca2019}, and {\tt matplotlib} \citep{Hunter:2007} python packages. The color palettes employed in this work are taken from \url{http://colorpalettes.net}.

    The Hyper Suprime-Cam (HSC) collaboration includes the astronomical communities of Japan and Taiwan, and Princeton University. The HSC instrumentation and software were developed by the National Astronomical Observatory of Japan (NAOJ), the Kavli Institute for the Physics and Mathematics of the Universe (Kavli IPMU), the University of Tokyo, the High Energy Accelerator Research Organization (KEK), the Academia Sinica Institute for Astronomy and Astrophysics in Taiwan (ASIAA), and Princeton University. Funding was contributed by the FIRST program from Japanese Cabinet Office, the Ministry of Education, Culture, Sports, Science and Technology (MEXT), the Japan Society for the Promotion of Science (JSPS), Japan Science and Technology Agency (JST), the Toray Science Foundation, NAOJ, Kavli IPMU, KEK, ASIAA, and Princeton University. 
    
    This paper makes use of software developed for the Large Synoptic Survey Telescope. We thank the LSST Project for making their code available as free software at \url{http://dm.lsst.org}.
    
    The Pan-STARRS1 Surveys (PS1) have been made possible through contributions of the Institute for Astronomy, the University of Hawaii, the Pan-STARRS Project Office, the Max-Planck Society and its participating institutes, the Max Planck Institute for Astronomy, Heidelberg and the Max Planck Institute for Extraterrestrial Physics, Garching, The Johns Hopkins University, Durham University, the University of Edinburgh, Queen’s University Belfast, the Harvard-Smithsonian Center for Astrophysics, the Las Cumbres Observatory Global Telescope Network Incorporated, the National Central University of Taiwan, the Space Telescope Science Institute, the National Aeronautics and Space Administration under Grant No. NNX08AR22G issued through the Planetary Science Division of the NASA Science Mission Directorate, the National Science Foundation under Grant No. AST-1238877, the University of Maryland, and Eotvos Lorand University (ELTE) and the Los Alamos National Laboratory.
    
    Based in part on data collected at the Subaru Telescope and retrieved from the HSC data archive system, which is operated by Subaru Telescope and Astronomy Data Center at National Astronomical Observatory of Japan.

    This project used public archival data from the Dark Energy Survey (DES). Funding for the DES Projects has been provided by the U.S. Department of Energy, the U.S. National Science Foundation, the Ministry of Science and Education of Spain, the Science and Technology FacilitiesCouncil of the United Kingdom, the Higher Education Funding Council for England, the National Center for Supercomputing Applications at the University of Illinois at Urbana-Champaign, the Kavli Institute of Cosmological Physics at the University of Chicago, the Center for Cosmology and Astro-Particle Physics at the Ohio State University, the Mitchell Institute for Fundamental Physics and Astronomy at Texas A\&M University, Financiadora de Estudos e Projetos, Funda{\c c}{\~a}o Carlos Chagas Filho de Amparo {\`a} Pesquisa do Estado do Rio de Janeiro, Conselho Nacional de Desenvolvimento Cient{\'i}fico e Tecnol{\'o}gico and the Minist{\'e}rio da Ci{\^e}ncia, Tecnologia e Inova{\c c}{\~a}o, the Deutsche Forschungsgemeinschaft, and the Collaborating Institutions in the Dark Energy Survey.
    
    The Collaborating Institutions are Argonne National Laboratory, the University of California at Santa Cruz, the University of Cambridge, Centro de Investigaciones Energ{\'e}ticas, Medioambientales y Tecnol{\'o}gicas-Madrid, the University of Chicago, University College London, the DES-Brazil Consortium, the University of Edinburgh, the Eidgen{\"o}ssische Technische Hochschule (ETH) Z{\"u}rich,  Fermi National Accelerator Laboratory, the University of Illinois at Urbana-Champaign, the Institut de Ci{\`e}ncies de l'Espai (IEEC/CSIC), the Institut de F{\'i}sica d'Altes Energies, Lawrence Berkeley National Laboratory, the Ludwig-Maximilians Universit{\"a}t M{\"u}nchen and the associated Excellence Cluster Universe, the University of Michigan, the National Optical Astronomy Observatory, the University of Nottingham, The Ohio State University, the OzDES Membership Consortium, the University of Pennsylvania, the University of Portsmouth, SLAC National Accelerator Laboratory, Stanford University, the University of Sussex, and Texas A\&M University.
    
    Based in part on observations at Cerro Tololo Inter-American Observatory, National Optical Astronomy Observatory, which is operated by the Association of Universities for Research in Astronomy (AURA) under a cooperative agreement with the National Science Foundation.

  \appendix
  \section{Pixelization effects}\label{app:pix}

    As discussed in Sec.~\ref{ssec:pcl.shear}, the fact that the cosmic shear signal is only sampled at the discrete positions of detected sources complicates the process of correcting for the effects of a finite pixel size. To illustrate this, consider the limit where each pixel contains a large ($>100$) number of sources with measured shapes homogeneously distributed within the pixel. In this case, the value of $\shear$ measured in each pixel corresponds to the average of the shear field within the pixel area, and therefore one must correct for the associated smoothing. In the opposite limit, in which all pixels contain either 0 or 1 sources, the shear value contained in each non-empty pixel corresponds to the value of the shear field sampled (as opposed to averaged) at the source positions, and no smoothing operation has taken place. Finite pixel effects therefore depend on the number density of the sample and the choice of pixel resolution.
    
\begin{figure}
      \centering
      \includegraphics[width=0.49\textwidth]{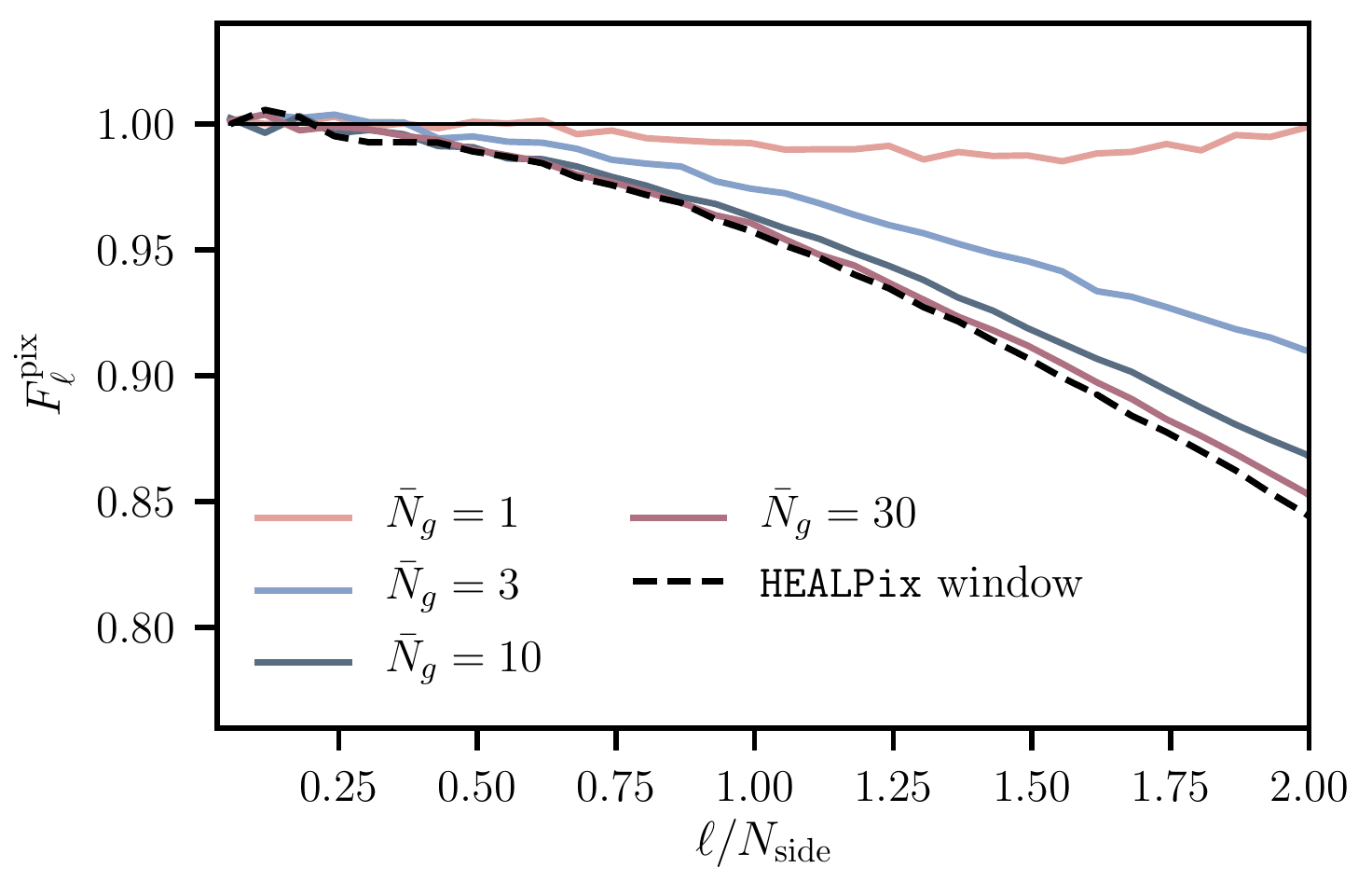}
      \includegraphics[width=0.49\textwidth]{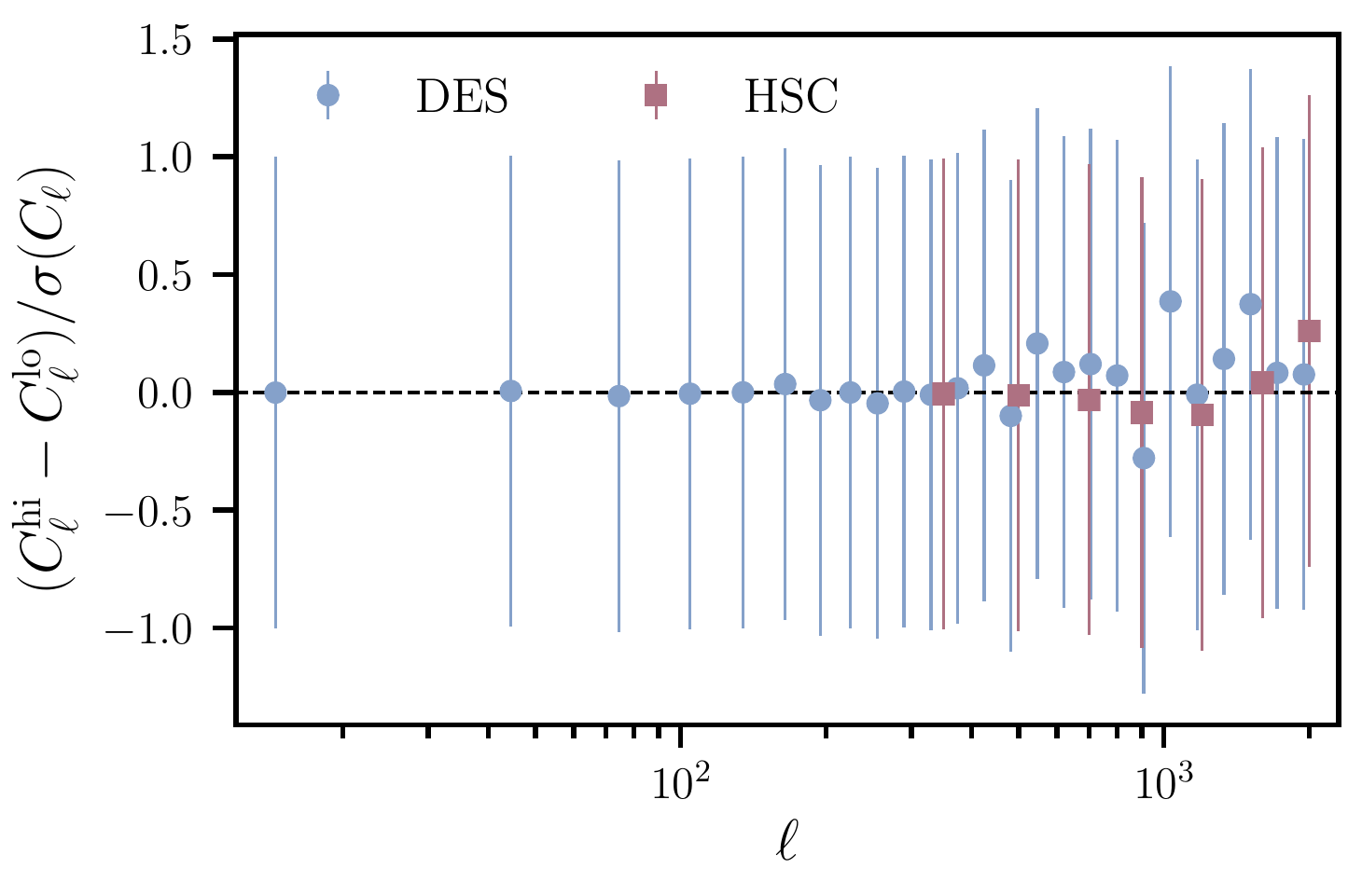}
      \caption{{\sl Left panel:} effective pixel window function as a function of the average number of shear sources per pixel $\bar{N}_g$. {\sl Right panel:} difference between the shear power spectrum computed with our fiducial pixel size and using pixels twice as large, shown as a fraction of their $1\sigma$ statistical uncertainties. Results are shown for the cross-correlation between the third and fourth redshift bins, which features the smallest error bars. The impact of the choice of pixelization is negligible within the range of scales used here.}
      \label{fig:pixwin}
    \end{figure}

    We can demonstrate this qualitatively through the following analysis. We start by generating a Gaussian realization of the shear field for the second DES redshift bin at a \texttt{HEALPix} resolution $N_{{\rm side},{\rm hi}}$. We then sample the values of this field at the positions of a number of randomly distributed points. These are then transformed into a shear map at a lower resolution $N_{{\rm side},{\rm lo}}<N_{{\rm side},{\rm hi}}$, and its power spectrum is computed using the methods described in Sec.~\ref{sec:pcl}. We repeat this for a total of 1000 realizations and then estimate an effective pixel window function $F_\ell^{\rm pix}$ as
    \begin{equation}
      F_\ell^{\rm pix} = \sqrt{\frac{\bar{C}_\ell}{C_\ell^{\rm th}}},
    \end{equation}
    where $\bar{C}_\ell$ is the measured $E$-mode power spectrum averaged over all simulations, and $C_\ell^{\rm th}$ is the input power spectrum. We repeat this for different numbers of randomly distributed points to study the dependence of the effective window function on the source number density. In order to limit computation time, we perform this analysis for $N_{{\rm side},{\rm hi}}=256$ and $N_{{\rm side},{\rm lo}}=64$. The results are shown in the left panel of Fig.~\ref{fig:pixwin}. As can be seen, we find that, in the small-density limit, the pixel window is close to 1 in the range of multipoles explored. For high number densities on the other hand, it approaches the theoretical pixel window function (shown as dashed black line), which corresponds to the case of exact averaging. We repeat this analysis for random points clustered assuming a linear bias model, finding no significant difference in the results.
    
    As stressed in Sec.~\ref{ssec:pcl.shear}, the recommended procedure is to use a pixel scale that is much smaller than the smallest scale used in the analysis, so these effects can be disregarded, and to verify that the measured power spectra are insensitive to the choice of pixel resolution, as we do in this work. The right panel of Fig.~\ref{fig:pixwin} shows the difference between the shear power spectra computed with the DES and HSC datasets using our fiducial pixelization and doubling the pixel size as a fraction of the statistical uncertainties. Within the range of scales relevant for cosmological analyses ($\ell\leq2000$), the effects of pixelization are negligible.

  \section{Null spectra}\label{app:nulls}
\begin{figure}[htb]
      \centering
      \includegraphics[width=\textwidth]{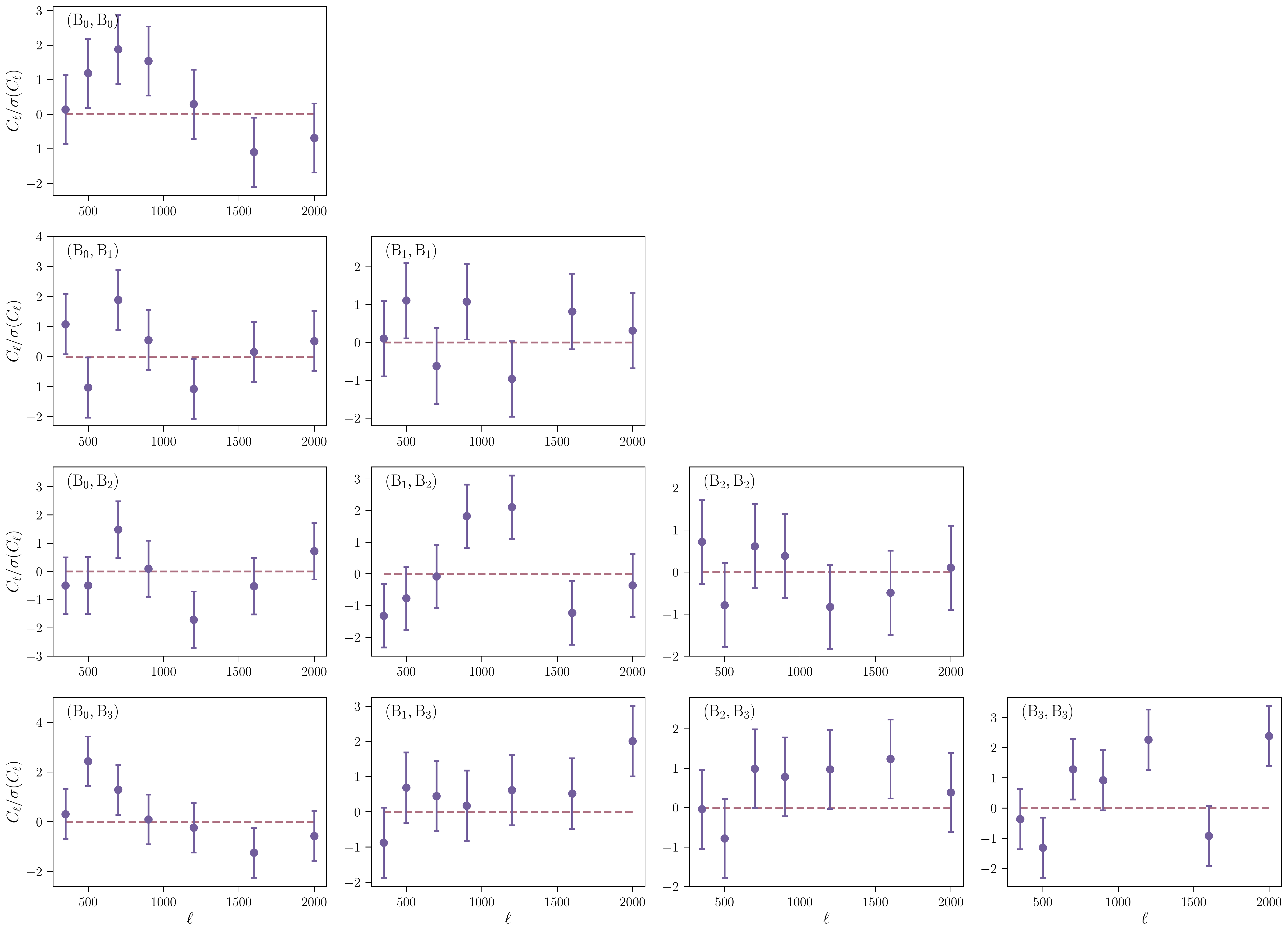}
      \caption{Angular $B$-mode power spectra for HSC, shown as a fraction of their $1\sigma$ statistical uncertainties.}
      \label{fig:hsc.Cls_bb}
    \end{figure}

\begin{figure}[htb]
      \centering
      \includegraphics[width=\textwidth]{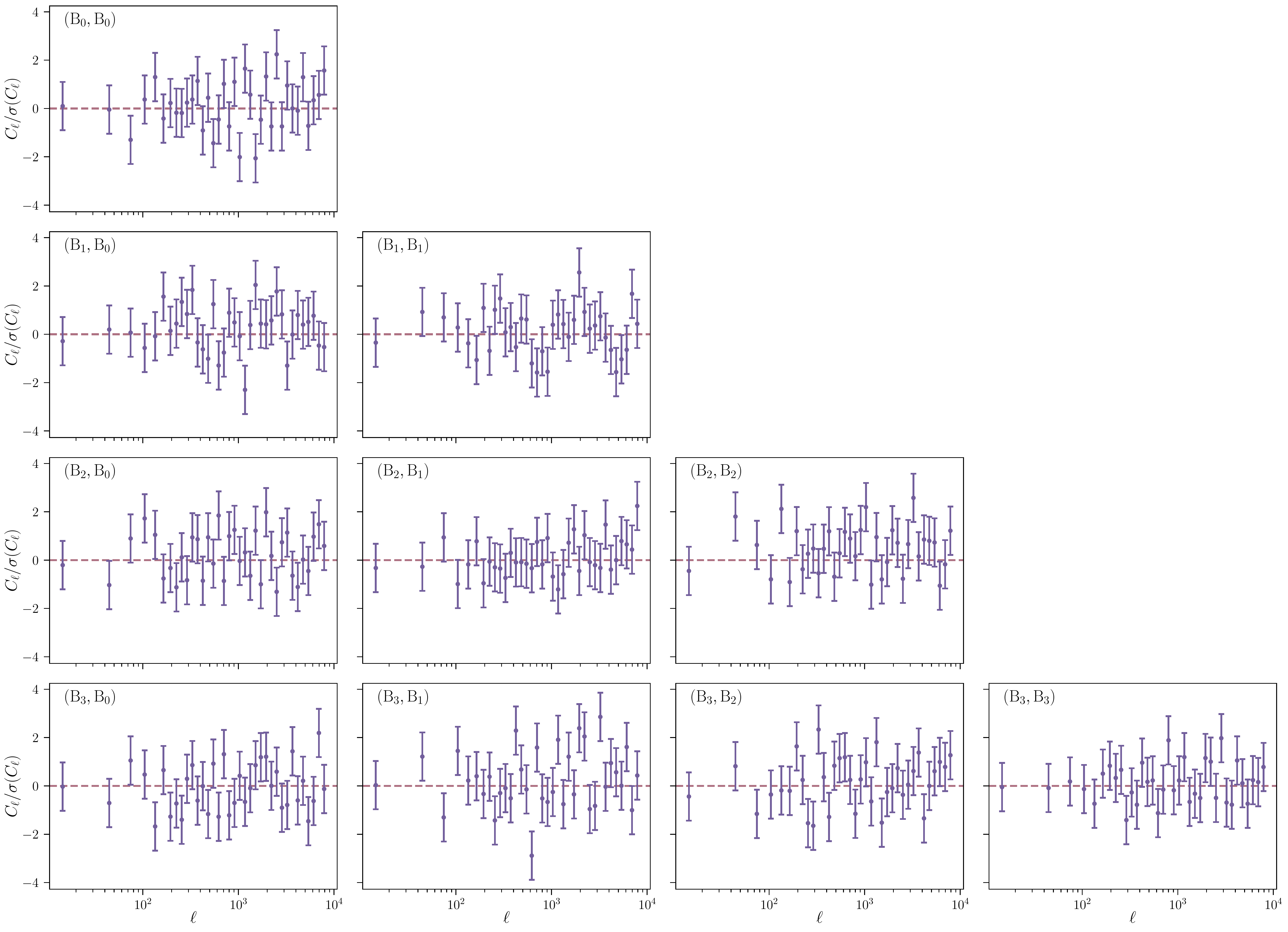}
      \caption{Angular $B$-mode power spectra for DES Y1, shown as a fraction of their $1\sigma$ statistical uncertainties.}
      \label{fig:des.Cls_bb}
    \end{figure}

\begin{figure}[htb]
      \centering
      \includegraphics[width=\textwidth]{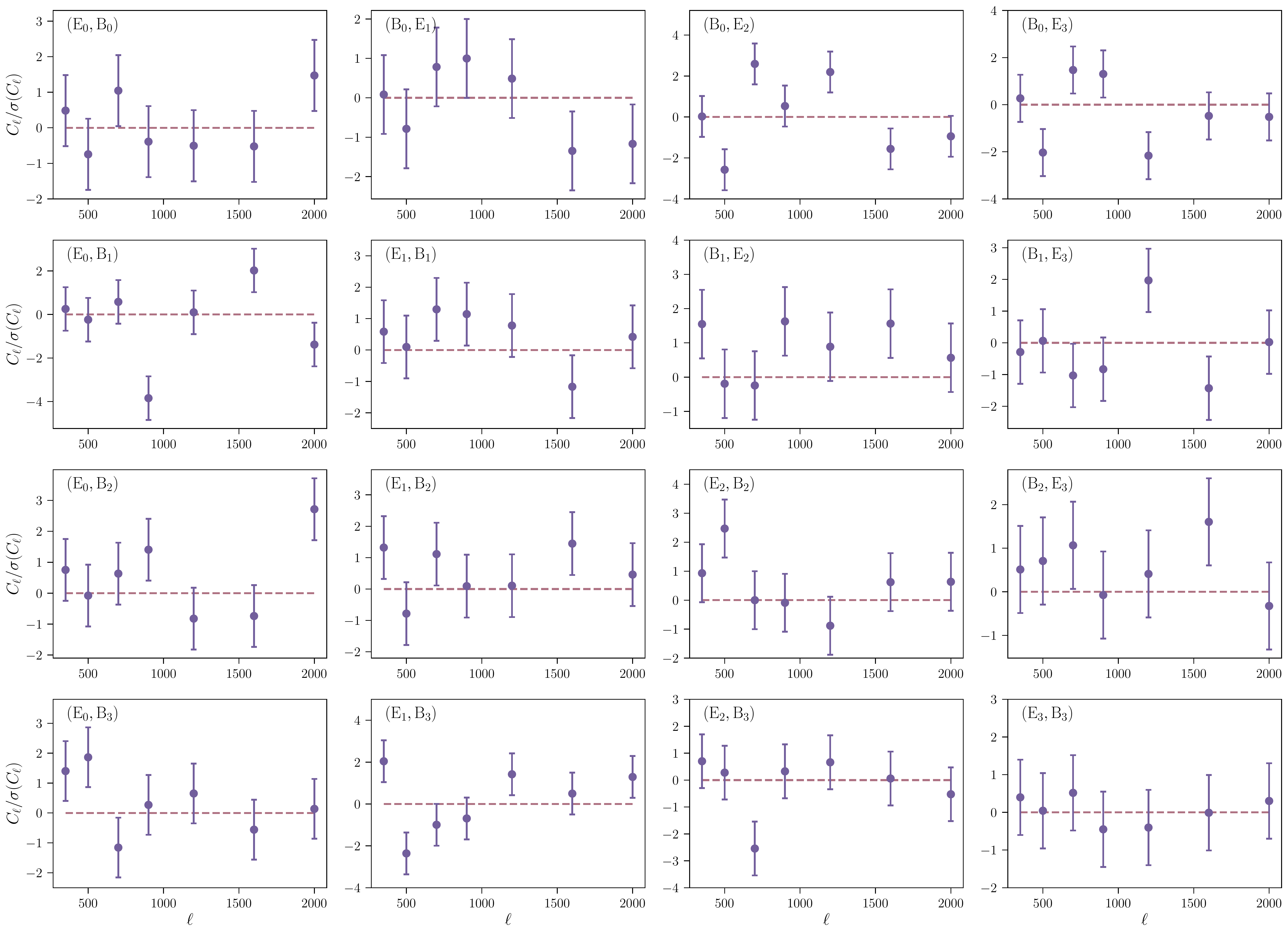}
      \caption{Cross-correlations between $E$ and $B$ shear modes for HSC, shown as a fraction of their $1\sigma$ statistical uncertainties.}
      \label{fig:hsc.Cls_eb}
    \end{figure}

\begin{figure}[htb]
      \centering
      \includegraphics[width=\textwidth]{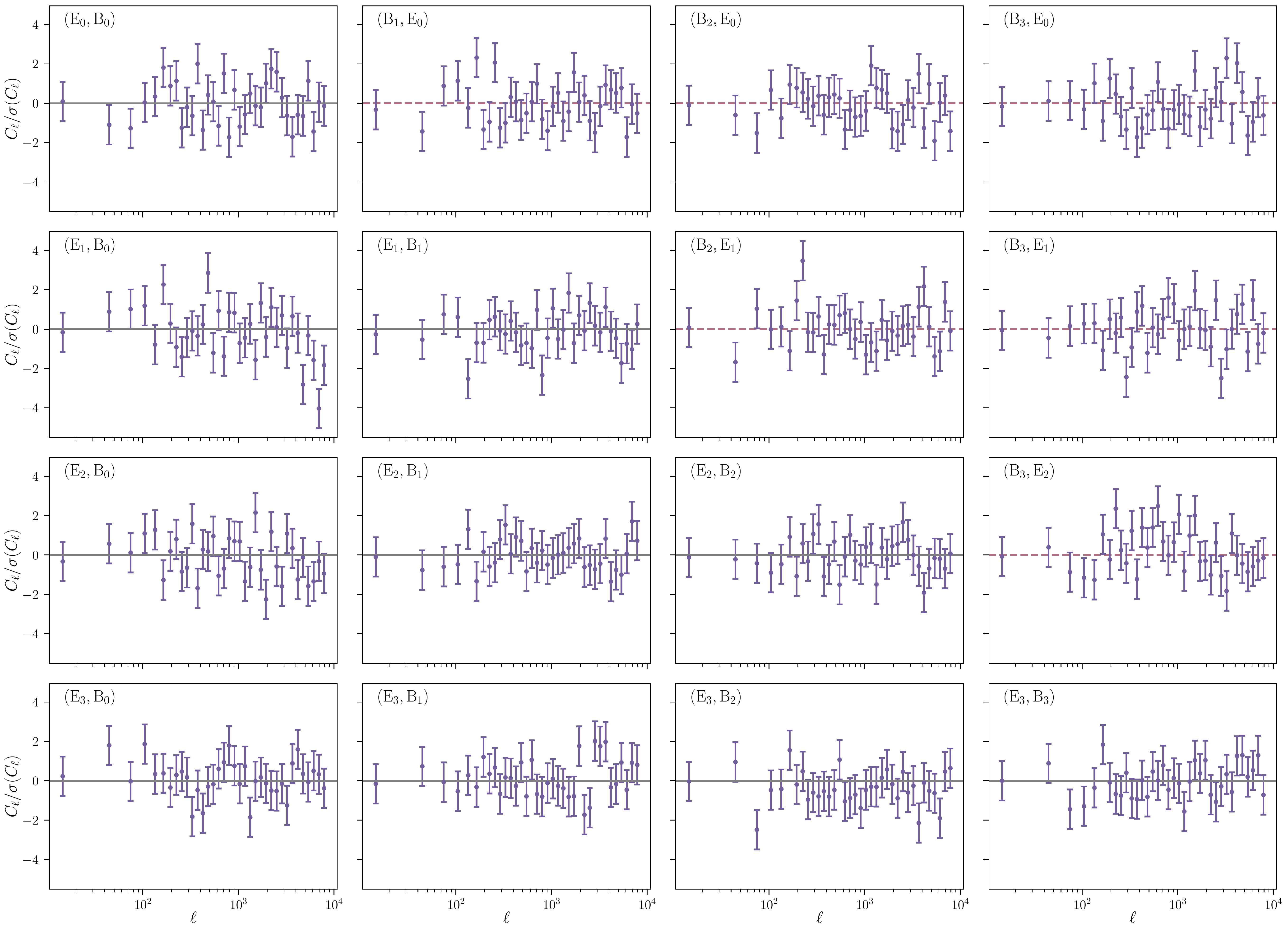}
      \caption{Cross-correlations between $E$ and $B$ shear modes for DES Y1, shown as a fraction of their $1\sigma$ statistical uncertainties.}
      \label{fig:des.Cls_eb}
    \end{figure}

\begin{figure}[htb]
      \centering
      \includegraphics[width=\textwidth]{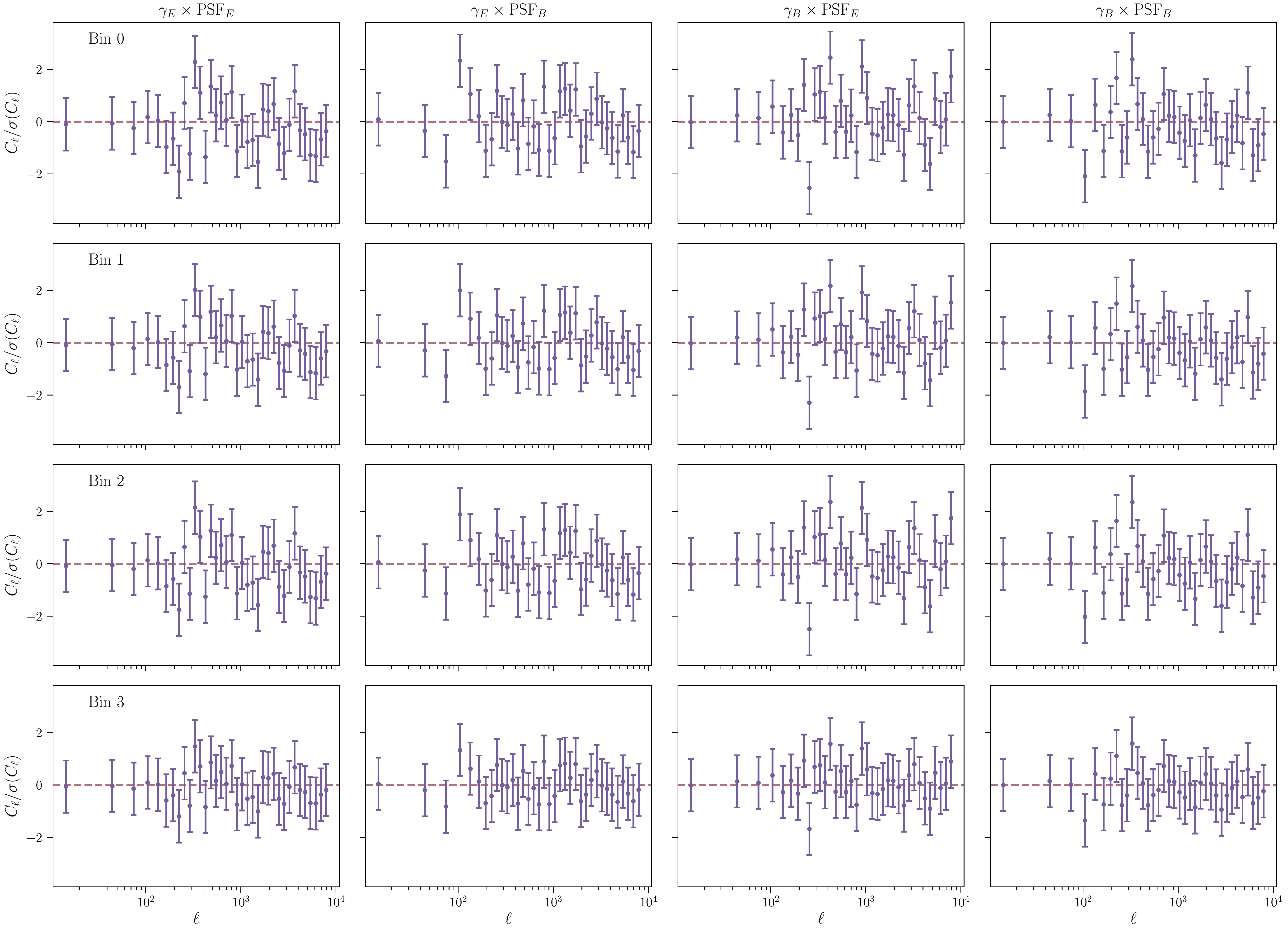}
      \caption{Angular power spectra of all possible cross-correlations between galaxy and PSF ellipticities for DES, shown as a fraction of their $1\sigma$ statistical uncertainties.}
      \label{fig:des.psf-ei}
    \end{figure}

\begin{figure}[htb]
      \centering
      \includegraphics[width=\textwidth]{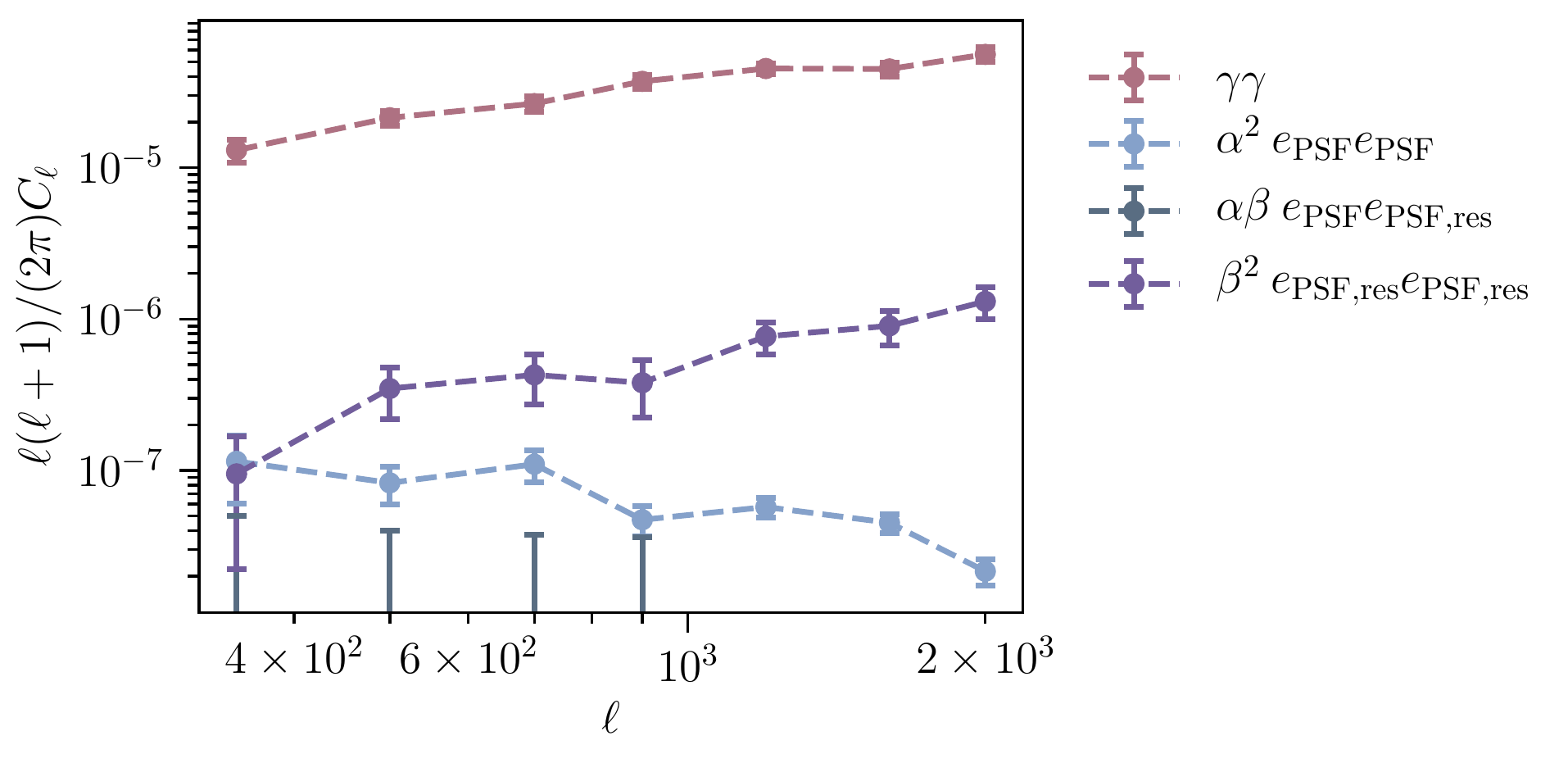}
      \caption{Comparison of the PSF leakage power spectrum and the PSF residual power spectrum to the non-tomographic cosmic shear power spectrum in HSC.}
      \label{fig:hsc.psf}
    \end{figure}

    The $BB$ power spectra of the cosmic shear field for the HSC and DES datasets are shown in Figures \ref{fig:hsc.Cls_bb} and \ref{fig:des.Cls_bb} respectively. The corresponding $EB$ spectra are shown in Figures \ref{fig:hsc.Cls_eb} and \ref{fig:des.Cls_eb}. The cross-spectra between the cosmic shear field and the PSF ellipticity maps for each tomographic bin are shown in Fig.~\ref{fig:des.psf-ei} for DES. For HSC, we show the comparison of the predicted PSF leakage power spectra to the non-tomographic cosmic shear signal in Fig.~\ref{fig:hsc.psf}. The quantitative description of these null tests is provided in Sec.~\ref{ssec:results.val}.

\bibliography{bibliography,non_ads}

\end{document}